\documentclass[12pt]{article}

\input{epsf}
\usepackage[dvips]{graphicx}
\usepackage{cite}
\usepackage{color}
\def\1ad{\mbox{\normalsize $^1$}}
\def\2ad{\mbox{\normalsize $^2$}}
\def\3ad{\mbox{\normalsize $^3$}}
\def\4ad{\mbox{\normalsize $^4$}}

\def\5ad{\mbox{\normalsize $^5$}}
\def\6ad{\mbox{\normalsize $^6$}}
\def\7ad{\mbox{\normalsize $^7$}}
\def\8ad{\mbox{\normalsize $^8$}}

\def\dj{\hbox{d\kern-0.347em \vrule width 0.3em height 1.252ex depth
-1.21ex \kern 0.051em}}

\newcommand{\be}{\begin{equation}}
\newcommand{\ee}{\end{equation}}
\newcommand{\ben}{\begin{equation*}}
\newcommand{\een}{\end{equation*}}
\newcommand{\ba}{\begin{eqnarray}}
\newcommand{\ea}{\end{eqnarray}}
\newcommand{\ban}{\begin{eqnarray*}}
\newcommand{\ean}{\end{eqnarray*}}
\newcommand{\brr}{\begin{array}}
\newcommand{\err}{\end{array}}
\newcommand{\bc}{\begin{center}}
\newcommand{\ec}{\end{center}}

\newcommand{\bea}{\begin{eqnarray}}
\newcommand{\eea}{\end{eqnarray}}
\newcommand{\bean}{\begin{eqnarray*}}
\newcommand{\eean}{\end{eqnarray*}}

\newcommand\lsim{\mathrel{\rlap{\lower4pt\hbox{\hskip1pt$\sim$}}
    \raise1pt\hbox{$<$}}}
\newcommand\gsim{\mathrel{\rlap{\lower4pt\hbox{\hskip1pt$\sim$}}
    \raise1pt\hbox{$>$}}}

\newcommand{\centeron}[2]{{\setbox0=\hbox{#1}\setbox1=\hbox{#2}\ifdim
                             \wd1>\wd0\kern.5\wd1\kern-.5\wd0\fi \copy0
                             \kern-.5\wd0\kern-.5\wd1\copy1\ifdim\wd0>\wd1
                             \kern.5\wd0\kern-.5\wd1\fi}}
\newcommand{\ltap}{\>\centeron{\raise.35ex\hbox{$<$}}
                     {\lower.65ex\hbox{$\sim$}}\>}
\newcommand{\gtap}{\>\centeron{\raise.35ex\hbox{$>$}}
                     {\lower.65ex\hbox{$\sim$}}\>}

\begin{document} 

\setcounter{page}{0}
\thispagestyle{empty}

\begin{flushright}
CERN-PH-TH/2007-007\\
hep-ph/0701158 \\
\today
\end{flushright}

\vskip 8pt

\begin{center}
{\bf \Large {
Multi-W Events at LHC from a \\ [0.25cm]  
Warped Extra Dimension with Custodial Symmetry
}}
\end{center}

\vskip 10pt

\begin{center}
{\large  Christopher Dennis$^{a}$, M\"uge Karag\"oz \"Unel$^{a}$,  \\ [0.25cm]  
 G\'eraldine Servant $^{b,c}$ and Jeff Tseng$^{a}$ }
\end{center}

\vskip 20pt
\begin{center}
\centerline{$^{a}${\it University of Oxford, Subdepartment of Particle Physics,
Keble Road, Oxford, OX1 3RH, UK}}
\centerline{$^{b}${\it CERN, Theory Division, CH-1211 Geneva 23, Switzerland}}
\centerline{$^{c}${\it Service de Physique Th\'eorique, CEA Saclay, F91191 Gif--sur--Yvette,
France}}
\vskip .3cm
\centerline{\tt C.Dennis2@physics.ox.ac.uk, karagozm@cern.ch}
\centerline{\tt geraldine.servant@cern.ch,j.tseng1@physics.ox.ac.uk}
\end{center}

\vskip 13pt

\begin{abstract}
\vskip 3pt
\noindent
Randall--Sundrum models based on $SU(2)_L\times SU(2)_R$ with custodial
symmetry are compelling frameworks for building alternative models of
electroweak symmetry breaking.  A particular feature of these models is the
likely presence of light Kaluza-Klein fermions related to the right-handed top
quark. These can be as light as a few hundred GeV and still compatible with EW
precision constraints.  In this article, we study the detectability of four-$W$ final states at the LHC,  which arise from the pair-production and $tW$ decay of light Kaluza-Klein bottom quarks as well as light Kaluza-Klein quarks carrying electric charge 5/3.

\end{abstract}

\vskip 13pt
\newpage
\section{Introduction}

In the last eight years,  extra-dimensional models have been suggested to solve
the gauge hierarchy problem. Among them, the Randall--Sundrum (RS) model
\cite{Randall:1999ee} is the most appealing, where the hierarchy between the
electroweak (EW) and the Planck scales arises from a warped higher dimensional
spacetime. Variants of the original set-up have matured over the years.
Eventually, all Standard Model (SM) fields except the Higgs  (to solve the
hierarchy problem, it is sufficient that just the Higgs --or alternative
dynamics responsible for electroweak symmetry breaking-- be localized at the
TeV brane) have been promoted to bulk fields rather than brane fields.  It has
been shown that EW precision constraints are much ameliorated if the EW gauge
symmetry in the 5-dimensional bulk is enlarged to $SU(2)_L \times SU(2)_R \times U(1)_{X}$
\cite{Agashe:2003zs}.  The AdS/CFT correspondence suggests that this model is
dual to a strongly coupled CFT Higgs sector \cite{Arkani-Hamed:2000ds}. Also,
the  $SU(2)_L \times SU(2)_R$ gauge symmetry in the RS bulk implies the
presence of a global custodial isospin symmetry of the CFT Higgs sector, thus
protecting EW observables from excessive new contributions
\cite{Agashe:2003zs}. This gauge structure in warped space has also been used
to construct Higgsless models of EW symmetry breaking \cite{Csaki:2003zu}. 

In this framework,  Kaluza-Klein (KK) excitations of gauge bosons of mass
$M_{KK} \sim 3 $~TeV are allowed (even lower, $\sim$ 500 GeV, for Higgsless models).  Generically, the mass spectrum of fermionic
KK modes depends on the boundary conditions (BC) imposed on the TeV and Planck
branes, localized at the boundaries of the slice of $AdS_5$. These are commonly
modelled by either Neumann ($+$) or Dirichlet ($-$) BC in orbifold
compactifications. Early studies in Randall-Sundrum background considered
Neumann boundary conditions only, in which case all KK masses have to be larger
than the gauge boson KK mass of 3 TeV and will be difficult to produce at the LHC.  In recent  models,
different boundary conditions have been imposed, leading to richer
possibilities for model building and to the potential presence of light KK
fermions, observable at colliders.  These fields have for instance Dirichlet BC on the
Planck brane and Neumann BC on the TeV brane. They are denoted  ($-+$) KK
fermions, and do not have zero modes.

The heaviness of the top quark is explained by the localization of the wave function of the top quark zero mode near the TeV brane, guaranteeing a large Yukawa coupling with the Higgs.
In the initial model of Ref.~\cite{Agashe:2003zs}, the Right-Handed (RH) top quark is included in a doublet of the $SU(2)_R$
symmetry. Its $b_R$ partner does not have a zero mode but its first KK excitation
is expected to be light.  This mode mixes  with the SM bottom quark and induces
large corrections to  $Z\rightarrow b \overline{b}$, so its mass has to be at
least $\sim$ 1.5 TeV \cite{Agashe:2003zs}. However, it has been recently
pointed out in Ref.~\cite{Agashe:2006at} that the custodial symmetry, together
with a discrete $L\leftrightarrow R$ symmetry  and alternative $SU(2)_R$ assignments for the top and bottom quarks, can protect the $Zb\overline{b}$
coupling and allow light masses for the KK $b_R$ as low as a few hundreds of
GeV. This mode is accompanied by  other light degenerate KK quarks (carrying electric charge $Q=2/3,-1/3,5/3$), that have been named {``the custodians"} \cite{Contino:2006qr}. They are likely to be the lightest KK states in these models and could be produced at the LHC. 

The interesting phenomenology of light  ($-+$) KK fermions  was pointed out in
Ref.~\cite{Agashe:2004ci} and its extended version Ref.~\cite{Agashe:2004bm}, where it
was emphasized that light KK fermions are expected as a consequence of the
heaviness of the top quark. More precisely, KK partners of the RH top should be
light.  Even though the study of  \cite{Agashe:2004ci,Agashe:2004bm} deals with
the embedding of  these  $SU(2)_L \times SU(2)_R \times U(1)$ Randall--Sundrum
models into a GUT\footnote{In this case, an extra $Z_3$ symmetry can be imposed
to protect proton stability, leading to the stability of a stable light KK
right-handed neutrino.  }, the statement remains pretty general: ($-+$) KK
partners belonging to the $t_R$ multiplet are expected to be light and can be
probed at colliders. The possibility of pair-producing these KK fermions, leading to multi $W$ events,  was
discussed. Most of these signatures were specific to the GUT model except for
the production of the KK $b_R$ (that we denote $\tilde{b}_R$), which is common
to a large class of models.  In this paper, we focus on the pair production
and $tW$ decay of $\tilde{b}_R$, which leads to a very unique $4W$ + $b\overline{b}$ final state.
This is a promising signature for probing RS models with custodial
symmetry\footnote{Collider signatures of this class of models are only starting
to be investigated, see for instance \cite{Agashe:2006wa}.}.
Moreover, while we will focus on the $\tilde{b}_R$ in this paper, it
should be noted that
our study applies as well to pair production of other custodians such as the KK $b_L$ and the  exotic KK quark with electric charge $Q=5/3$, that both decay into $tW$.
 
 The pair production of exotic quarks, singlet under $SU(2)_L$, was considered in the past in Ref.~\cite{Aguilar-Saavedra:2005pv}. The single and pair production of heavy $T$ quarks in  Little Higgs models was studied in Ref.~\cite{Azuelos:2004dm}.
 However, these quarks have
charge +2/3 instead of -1/3 or 5/3, so the decay is into $W b$ rather than $W t$ and
there are only 2 $W$'s in the event. We are not aware of previous studies
investigating events with  more than 2 $W$'s in the final state\footnote{Multiple weak gauge boson production was theoretically investigated in the eighties \cite{Chanowitz:1984ne,Golden:1985qg} and early nineties \cite{Morris:1992nr,Barger:1989cp,Barger:1991jn} in the context of the Superconducting Super Collider (SSC), as these processes were (and are) of great interest to test whether the EW symmetry breaking is due to new strong interactions. If this is the case, multiple ($>2$) longitudinal $W$ and $Z$ production would be enhanced  at SSC energies \cite{Chanowitz:1984ne,Morris:1992nr}. This is nevertheless still a small effect compared to the rates for the standard production of multiple transverse $W$ and $Z$ and it was concluded that it would not be observable \cite{Golden:1985qg}.
More recent work has suggested that these studies could be performed with
${\cal O}(1000)\;{\rm fb}^{-1}$ of data that could be delivered by an
upgraded LHC \cite{Gianotti:2005}.}.
While 4-$W$ production brings the prospect of a spectacularly rich multi-jet plus multi-lepton signature,
distinguishing the signal from backgrounds will still be challenging.
Our goal in this paper is to investigate the
feasibility  of doing so and thus identifying $\tilde{b}_R$ at the LHC.
  
 A straightforward trigger criterion for these events is that of a single,
isolated lepton with missing $E_T$, {\em i.e.}, the standard leptonic $W$ data
stream.  Other $W$'s in the event would be reconstructed using dijet pairs.
Multi-lepton triggers can also be used.  The fermions $\tilde{b}_R$ are
particularly accessible, producing $b$ jets which could be tagged with
displaced tracks.  If one $\tilde{b}_R$ enters the leptonic $W$ data stream,
its antiparticle partner is potentially amenable to full reconstruction.

In the present work, we neglect hadronization effects.  For the range of
masses considered, the
width of $\tilde{b}_R$ is large enough that the decay takes place before
hadronization.  

For low $\tilde{b}_R$ masses ($<300$ GeV), there are interconnection effects.  
This situation is common to nearly all investigations of heavy particles with
GeV-scale widths such as $WW$, $ZZ$, and $t \overline{t}$
where hadronic decays
overlap.
We assume this effect
is small; this is justified for the
${\cal O}(500)\;{\rm GeV}$ masses we consider in detail, and we neglect it for now
(it is interesting to note that on the opposite extreme, 
the KK fermions of \cite{Agashe:2004bm} are very
long-lived and lead to CHAMP-like signatures;
for these, we are well into the
realm of Heavy Quark Effective Theory, so that to first order  we can again
neglect light quark effects in the decay).

In Section \ref{sec:parameters}, we describe the parameters of the model and the decay channels of $\tilde{b}_R$. Readers who are more keen on the experimental aspects than on the model-building details can skip this section.
Section  \ref{eventselection} discusses the possible signatures associated with  $\tilde{b}_R$ pair production and presents the main SM backgrounds mimicking the $4W$ signature.
Section~\ref{sec:signal} gathers the results of our simulation and outlines a promising strategy for distinguishing the signal from the SM backgrounds
using early ($10\;{\rm fb}^{-1}$) LHC data.

\section{Model parameters}
\label{sec:parameters}

Our analysis describes the pair production and decay of $\tilde{b}_R$. However, as we said in the introduction, there are typically other ``custodian" quarks which lead to the same $4W$ signature, potentially significantly enlarging our signal.
The number of custodians is model-dependent. Following \cite{Agashe:2006at}, several papers have appeared recently on the phenomenology associated with these light KK fermions \cite{Contino:2006qr,Carena:2006bn,Djouadi:2006rk,Cacciapaglia:2006mz}. These studies consider different embeddings for the SM top and bottom quarks, leading to various possibilities for the number and masses of the custodians.
In the description below, we assume for simplicity that there is only one light KK quark, $\tilde{b}_R$, the $SU(2)_R$ doublet partner of the SM RH top quark. This is not the realistic situation but it does not matter for our analysis. At the end of this section, we comment how to generalize our discussion to  the more realistic models mentioned above.  

KK fermions are 4-component spinors which have both a LH and a RH chirality.
One chirality has ($-+$) BC while the other has ($+-$).  The RH chirality of the KK
$b_R$ under consideration turns out to be localized near the TeV brane and the
LH one near the Planck brane. Therefore, the couplings of the LH chirality with
modes localized near the TeV brane, such as the Higgs, the top quark and KK
excitations, are suppressed.  In contrast, the direct interactions of  KK
fermions to zero-mode gauge bosons are vector-like.  This is because zero-mode
gauge bosons have a flat profile (unlike KK modes) and couple identically to
both chiralities of KK fermions.  The couplings of the KK ${b}_R$ to SM gauge
bosons are the same as the couplings of the SM $b_R$ quark, except that they
involve both chiralities identically.  The vertices $\overline{\tilde{b}}_R
\tilde{b}_R G_{\mu}$, $\overline{\tilde{b}}_R \tilde{b}_R A_{\mu}$ and
$\overline{\tilde{b}}_R \tilde{b}_R Z_{\mu}$ are respectively
$g_s\gamma_{\mu}$, $-{e}\gamma_{\mu}/{3}$ and
${e\tan\theta_W}\gamma_{\mu}/{3}$.  $\tilde{b}_R$ has four decay channels:
$\tilde{b}_R \rightarrow t_R W$, $\tilde{b}_R \rightarrow t_L W$,  $\tilde{b}_R
\rightarrow b_L H$ and $\tilde{b}_R \rightarrow b_L Z $. The first two lead to
the multi $W$ signature we are interested in, while the third
can also lead to it
if the Higgs decays dominantly into $WW$.  We now present these decays in
details.

\begin{figure}[!htb]
\begin{center}
\includegraphics[height=3.75cm,width=5.5cm]{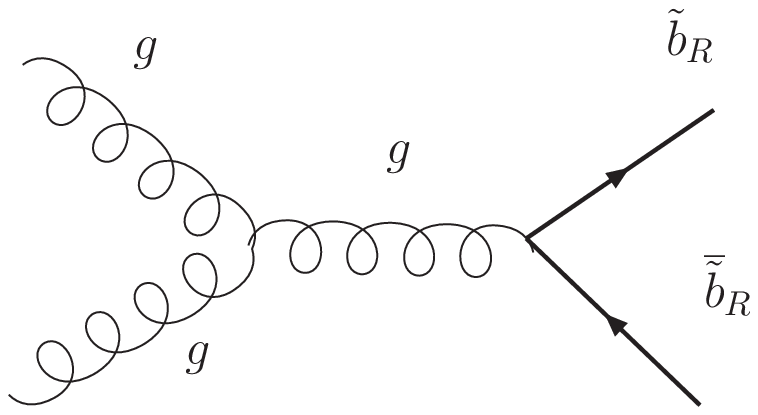}
\includegraphics[height=3.75cm,width=5.5cm]{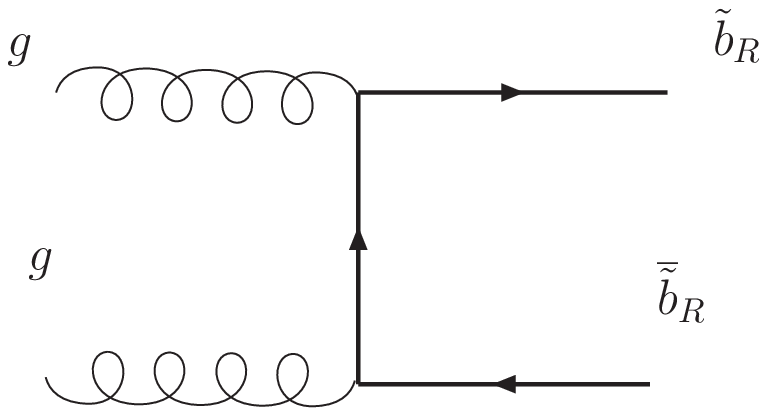}
\includegraphics[height=3.75cm,width=5.5cm]{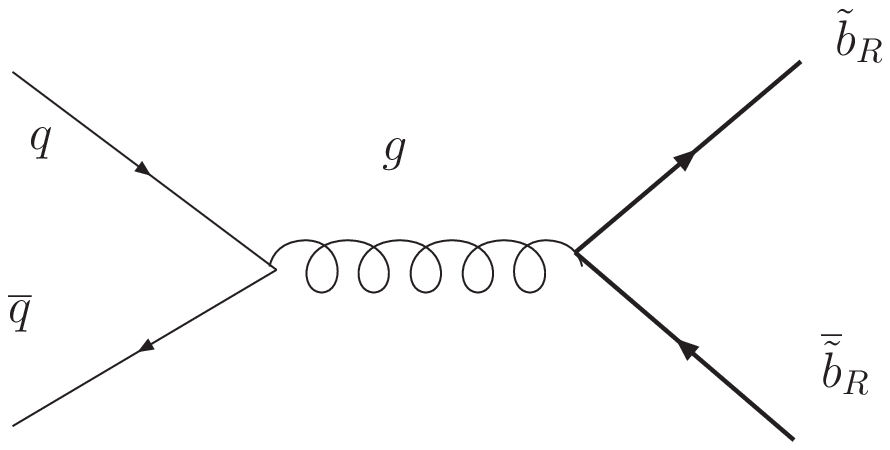}
\caption{Dominant channels for pair production of $\tilde{b}_R$. }
 \label{Production_channels}
\end{center}
\end{figure}
\begin{figure}[!htb]
\begin{center}
\includegraphics[height=3.75cm,width=4.cm]{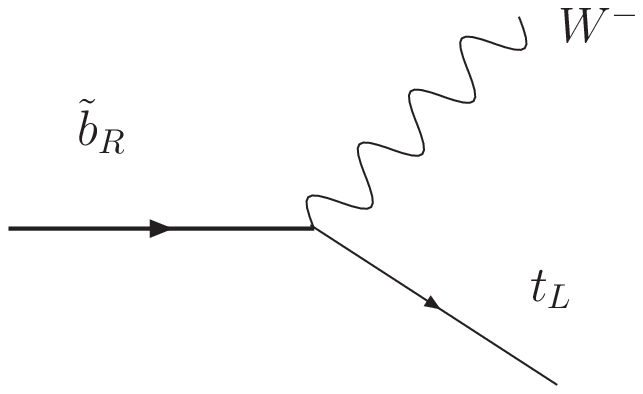}
\includegraphics[height=3.75cm,width=4.cm]{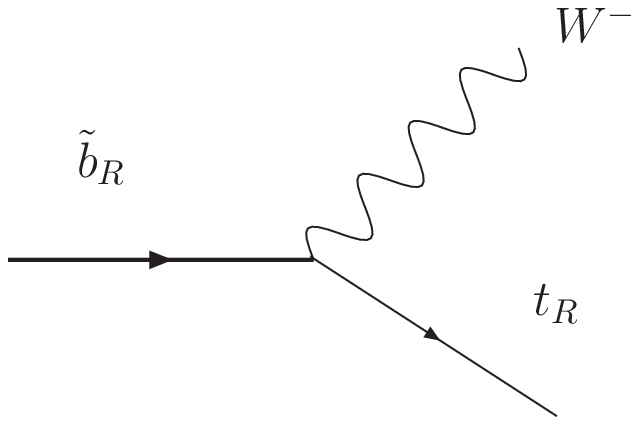}
\includegraphics[height=3.75cm,width=4.cm]{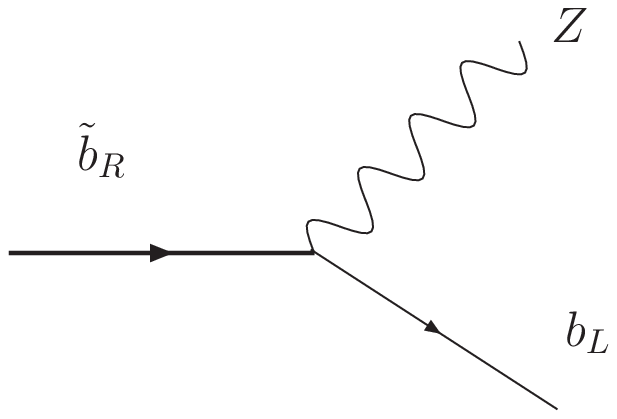}
\includegraphics[height=3.75cm,width=4.cm]{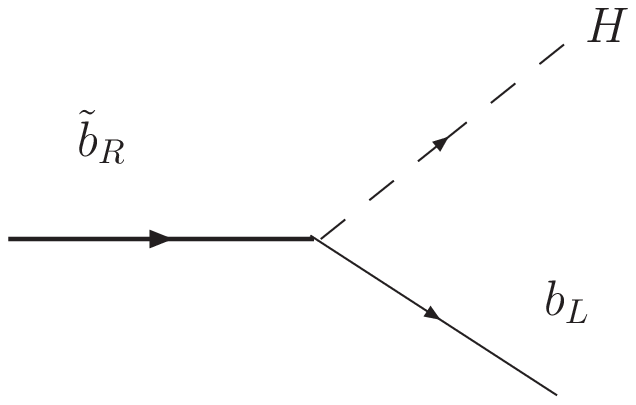}
\caption{The four decay channels of $\tilde{b}_R$. }
 \label{Decay_channels}
\end{center}
\end{figure}

\subsection{  $\tilde{b}_R \rightarrow t_R  W^-$ decay}

\begin{figure}[h]
\begin{center}
\includegraphics[width=4.5cm,height=2.5cm]{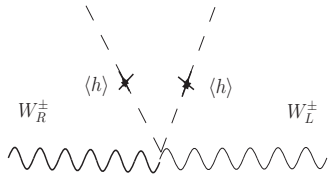}
\caption[]{Diagram leading to $W_R^{\pm}-W_L^{\pm}$ mixing.}
\label{fig:mixing}
\end{center}
\end{figure}

\begin{figure}[!htb]
\begin{center}
\includegraphics[height=6.5cm,width=10.3cm]{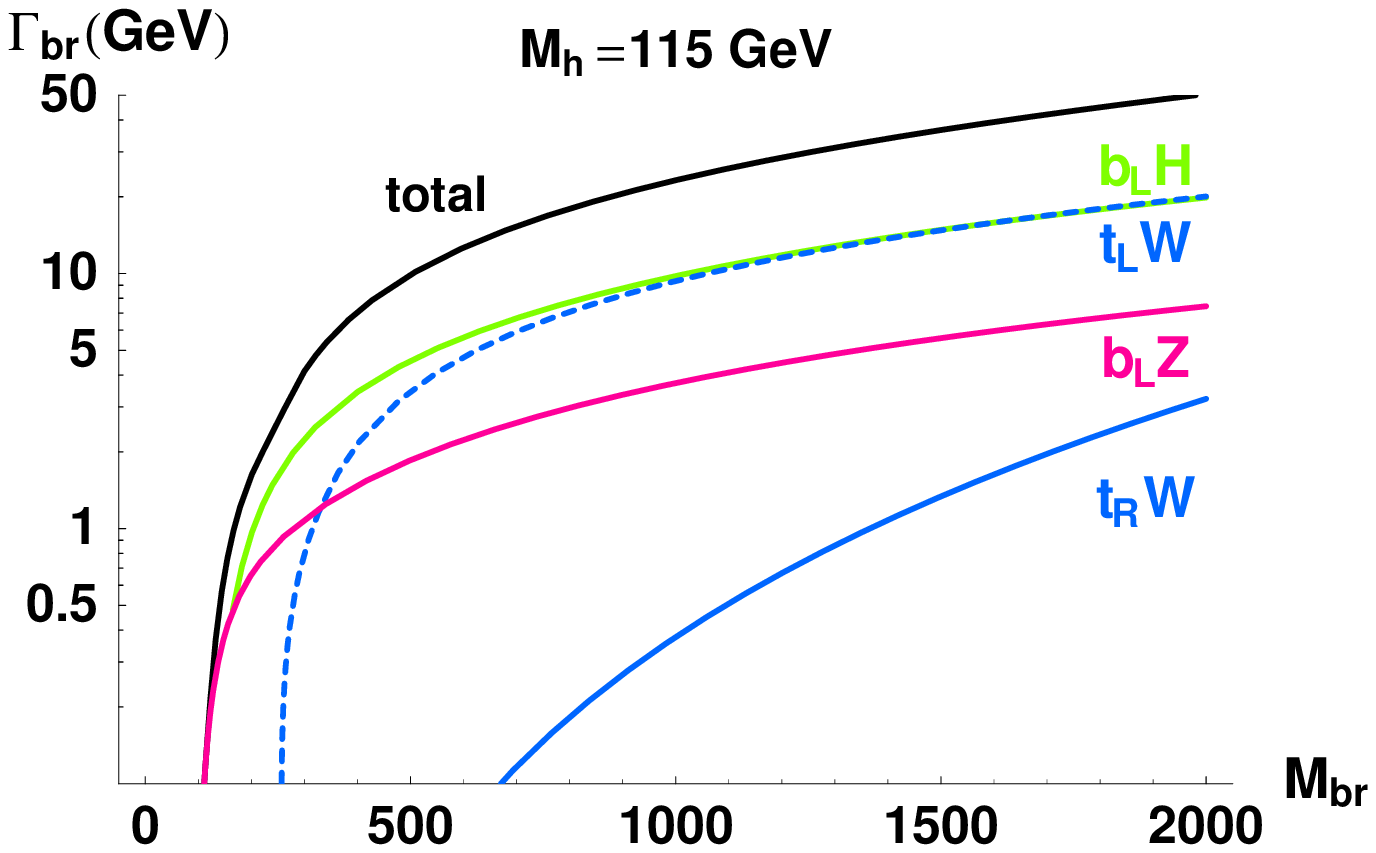}
\includegraphics[height=4.5cm,width=8.3cm]{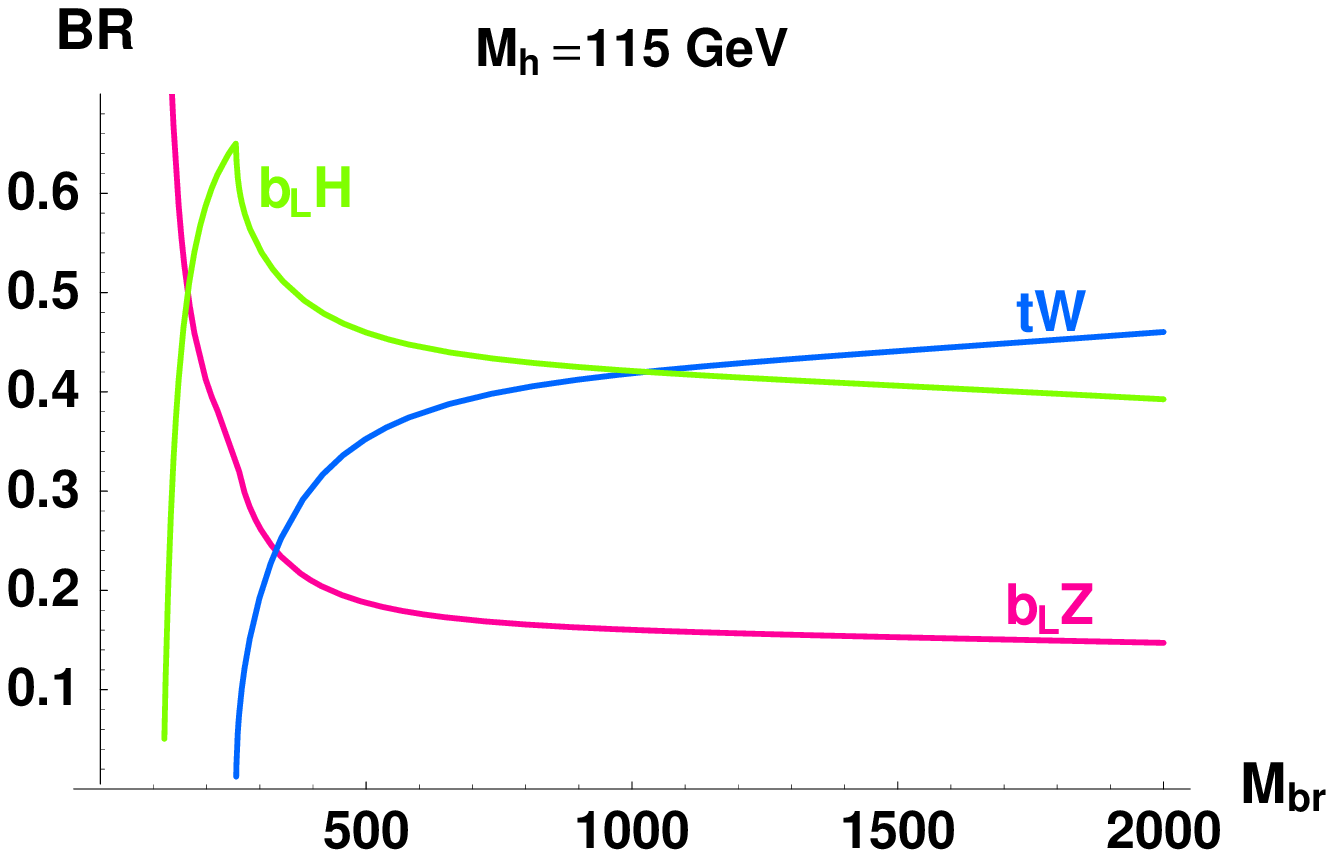}
\includegraphics[height=4.5cm,width=8.3cm]{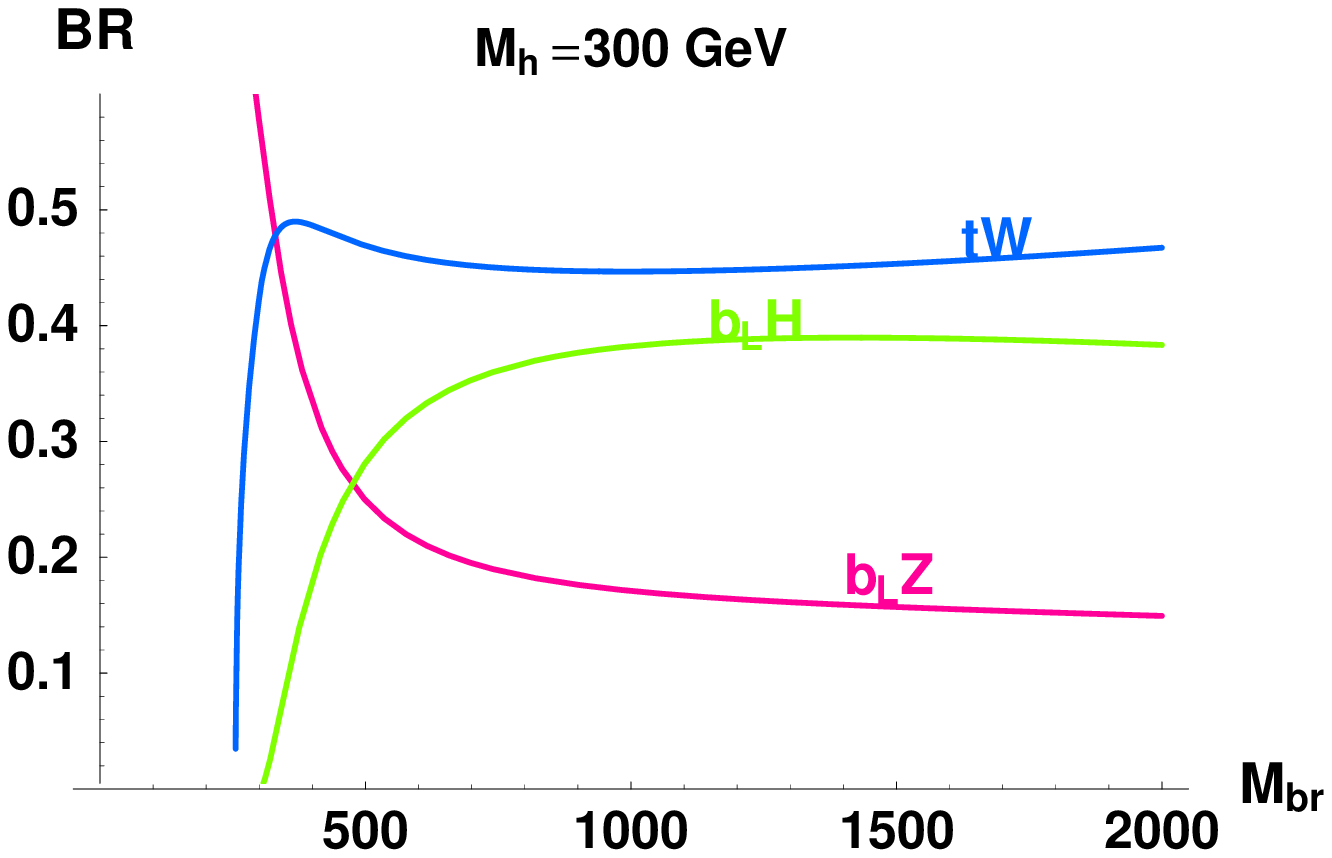}
\caption{Decay widths of $\tilde{b}_R$ as a function of its mass and
corresponding branching ratios.} \label{Decay_width}
\end{center}
\end{figure}

$\tilde{b}_R$, as a singlet under $SU(2)_L$, is not expected to couple to $W$.
The effective coupling of $\tilde{b}_R$ to $W$ is due to $W_R-W_L$ mixing
resulting from EW symmetry breaking (see Fig.~\ref{fig:mixing}).  $W_R$  is the
charged gauge boson associated with $SU(2)_R$ and does not have a zero mode.
We consider the effect of the mixing  of the SM $W$ with the first KK
excitation of $W_R$, which has mass $M_{KK}\sim 3$ TeV. The induced coupling is
then $g_{ \tilde{b}_R , t_R , W^-} \ \gamma_{\mu}$ where:
\begin{eqnarray}
g_{\tilde{b}_R , t_R , W^-}&= &
  \frac{g_{R}}{\sqrt{2}}\sqrt{k \pi r_c} \times P_R
  \times {\cal M}_{W_R - W_L}  \times {{\cal F}_{ \tilde{b}_R , t_R}}\\
{\cal M}_{W_R - W_L} &=&
  \frac{g_{R}}{g}\sqrt{2 k \pi r_c} \frac{M_W^2}{M_{KK}^2}.
\end{eqnarray}
${{\cal F}_{ \tilde{b}_R , t_R}}\sim 1$ is the form factor reflecting the
overlap between the wave functions of $\tilde{b}_R$, $W_R$ and $t_R$.  ${\cal
M}_{W_R - W_L} $ is the mixing factor due to the EW breaking vev of the Higgs.
$k \pi r_c=\ln (10^{15})$ is the exponent of the warp factor needed to generate
the weak/Planck hierarchy.  This formula assumes the Higgs is localized on the
TeV brane. If it is delocalized in the extra dimension as in scenarios of
gauge-Higgs unification where it is identified with the 5th component of a
gauge boson, we can replace the factor $\sqrt{2 k \pi r_c}$ with the form
factor accounting for the wave function overlap. In this case, using  the profile given in
\cite{Contino:2003ve} (see also Appendix B of \cite{Agashe:2004bm}) $\sqrt{2 k
\pi r_c}\approx 8.31$ is replaced by 5.748.  $P_R$ is the RH projector that
expresses the fact that only one chirality of the KK $b_R$ has a non-suppressed
coupling to $W_R$.  $g_R$ is the 4D gauge coupling of $SU(2)_R$.  The LR
discrete symmetry requires $g_L = g_R$ at 5D level. Since the 5D $g_L$ is fixed
by the matching with the 4D $g_L$ (assuming that brane kinetic terms are
small), we can replace $g_R$ by $g$, the  SM $SU(2)$ coupling, in the above
formulae.  As is clear from Fig.~\ref{Decay_width}, this decay is negligible
compared to the other three channels.

\subsection{$\tilde{b}_R \rightarrow t_L  W^-$,  $\tilde{b}_R \rightarrow b_L
Z$  and $\tilde{b}_R \rightarrow b_L  H$ decays}

The Yukawa coupling between the $ ( t_L,b_L )$ and  $( t_R, \tilde{b}_R)$
multiplets generates three more decay modes for  $\tilde{b}_R$: $\tilde{b}_R
\rightarrow b_L  H$, $\tilde{b}_R \rightarrow t_L  W^-$ and $\tilde{b}_R
\rightarrow b_L  Z$.

The coupling between $b_L$, $H$  and $\tilde{b}_R$ is $y_t
f(c_{t_R})/\sqrt{2}$, where $y_t=1$ and $f (c ) \approx \sqrt{ 2 / ( 1 - 2 c )
}$ (for $c \gsim -1/2$) comes from the wavefunction of the RH top quark given in
appendix A of \cite{Agashe:2004bm}.  The $c$-parameter corresponds to the 5D
fermionic bulk mass in Planck units. For the right-handed top quark, we take
$c=-1/2$, leading to $f(c _{t_R})=1$.

The second decay comes from the interaction $\tilde{b}_R . t_L . H^{\pm}$ where
$H^{\pm}$ is the would-be Goldstone boson which becomes a longitudinal $W$.  In
unitary gauge, there is no $H^{\pm}$, and this has to be written in the form of
a $\gamma_{\mu}$ interaction.  However, such an
interaction will couple a LH (RH)
fermion with a LH (RH) fermion.  As we said earlier, the RH chirality of the KK
$b_R$ under consideration turns out to be localized near the TeV brane and the
LH one near the Planck brane. Therefore, the coupling of the LH chirality with
the top quark is suppressed (since the top is localized near the TeV brane).
One would be tempted to conclude that there is no coupling between
$\tilde{b}_R$ $t_L$
and $W$, but there is a subtlety: the LH chirality of the {\it physical}
(mass eigenstate) ${b} _R$ is actually a linear combination of a LH KK $b_R$
and a zero mode $b_L$.  There is some mixing effect (comparable to the one
discussed in section 9.3 of \cite{Agashe:2004bm}) due to the top Yukawa
coupling which generates a mass term between $b_L$ and $\tilde{b}_R$.  After
diagonalization of the mass matrix, we can see that the new mass eigenstate
$\tilde{b}_R$ has an admixture of the zero mode $b_L$, through which it couples to
$W$ and $Z$.  Specifically, in unitary gauge, 
\be
[ \mbox{physical} \  \tilde{b} ]_{LH} =
  \cos \theta \ \hat{ \tilde{b}}_R  + \sin \theta \ b_L
\ee
where $ \hat{ \tilde{b}}_R  $  denotes the LH chirality of KK $b_R$, and
$\sin \theta \approx m_t  f(c _{t_R}) / m_{ \tilde{b}_R }$ comes from the mass
term between $b_L$ and $\tilde{b}_R$.  Via the $b_L$ component, the physical
$\tilde{b} $ couples to $W$ and $t_L$ (with SM coupling $g/\sqrt{2}$). When we
project onto the longitudinal $W$, we get a factor of $E / m_W$ from the
polarization vector, where $E \sim m_{ \tilde{b}_R }$ since we are considering
decays of $\tilde{b}_R$. The coupling to $W_{long}$ is $y_{t} f(c _{t_R})$ as
expected using the goldstone equivalence theorem (coupling to charged Higgs).
To summarize, in unitary gauge, the { physical}  $\tilde{b}_R$ has  the
following interactions: \\
$\tilde{b}_R  ( g_{Zb_L}  b_L  Z + (g/\sqrt{2}) \  t_L  W )  m_t 
  f(c _{t_R}) / m_{\tilde{b}_R}$ where $g_{Zb_L}=
 (g/\cos \theta_W)(1/2 -\sin^2\theta_W/3)$.

\subsection{Decay widths}
\label{subsec:widths}
The partial width for the decay $\tilde{b}_R \rightarrow t_R  W^-$ is
\bea
\Gamma_{ t_R  W^-}= \ \frac{{g_{Wtr}}^2 \left[{m_{\tilde{b}_R}}^4+({m_W}^2-2 {m_t}^2)
   {m_{\tilde{b}_R}}^2+{m_t}^4-2 {m_W}^4+{m_t}^2 {m_W}^2\right]}
   {32\pi {m_{\tilde{b}_R}} {m_W}^2  } \ 
 \lambda^{1/2}(m_t,m_W,m_{\tilde{b}_R})
\eea
where 
\bea
g_{Wtr}=g_R \ k \pi r_c \left(\frac{m_W}{M_{KK}}\right)^2 \ \ \mbox{and} \ \
 \lambda(m_t,m_W,m_{\tilde{b}_R})=   1+ \frac{(m_t^2-m_W^2)^2}{m^4_{\tilde{b}_R}}-2\frac{(m_t^2+m_W^2)}{m^2_{\tilde{b}_R}} .
\eea
We use the same formula for the decay $\tilde{b}_R \rightarrow t_L  W^-$ except that the coupling constant is replaced by 
\be
g_{Wtl}=f(c_{t_R}) \frac{m_t}{m_{\tilde{b}_R}} \times \frac{g}{\sqrt{2}} .
\ee
The decay width for  $\tilde{b}_R \rightarrow b_L  Z$ is
\be
\Gamma_{ b_L  Z}=\frac{{g_{Zbl}}^2 {\left({m_{\tilde{b}_R}}^2-{m_Z}^2\right)}
   \left({m_{\tilde{b}_R}}^4+{m_Z}^2 {m_{\tilde{b}_R}}^2-2 {m_Z}^4\right)}{32\pi \  {m^3_{\tilde{b}_R}}
   {m_Z}^2  }
\ee
where
\be
g_{Zbl}=f(c_{t_R}) \frac{m_t}{m_{\tilde{b}_R}} \times \frac{g}{6 \cos \theta_W} [3-2\sin^2 \theta_W],
\ee
and finally
\be
\Gamma_{ b_L  H}=\frac{f^2(c_{t_R}) \left({m_{\tilde{b}_R}}^2-{m_H}^2\right)^2}{32 \pi {m^3_{\tilde{b}_R}} }.
\ee
As we said, in our analysis, we use $f(c_{t_R})=1$, $g_R=g$, $M_{KK}=3$ TeV,
and $k\pi r_c=\ln(10^{15})$, so that the only free parameter is $m_{\tilde{b}_R}$.

In our formulae, we have assumed that $(t_R,\tilde{b}_R)$ form an $SU(2)_R$ doublet even though we know that this is not the realistic situation. The solution to the $Zb\overline{b}$ problem indeed requires that $t_R$ is either a singlet under $SU(2)_R$ or belongs to a
$({\bf{1,3}})+({\bf 3,1})$  under $SU(2)_L\times SU(2)_R$ \cite{Agashe:2006at}. In the first case, there is obviously no $\tilde{b}_R$. However, this does not mean that there is no associated light KK quark. Indeed, in gauge-Higgs unification models (see Ref.~\cite{Contino:2003ve}) which are presumably the best motivated models for EW symmetry breaking in RS, the SM $t_R$ belongs to a larger multiplet which necessarily leads to light custodian partners. In minimal composite Higgs models, $t_R$ belongs to a {\bf 5}=(({\bf 2,2}),({\bf 1,1})) or a {\bf 10}=(({\bf 2,2}),({\bf 1,3})+({\bf 3,1})) of $SO(5)$. Its $({\bf 2,2})$ partners contain the SM $Q_L$ as well as an extra doublet of custodians of electric charge (2/3,5/3). The Q=2/3 quark will not lead to the $4W$ signature, but the Q=5/3 quark  (that we name $\tilde{q}$) will, and with a branching ratio essentially equal to 1. If $t_R$ belongs to a {\bf 10}, there will be as well two bottom-like custodians $\tilde{b}_R$ and $\tilde{b}_L$, and  two alike $\tilde{q}$ custodians all leading to the $4W$ signature. All these particles have the same mass, so they are produced with the same cross section. The  $L\leftrightarrow R$ discrete symmetry also guarantees the same couplings for $\tilde{b}_R$ and $\tilde{b}_L$ and for $\tilde{q}_R$ and $\tilde{q}_L$.
However,  the fact that they belong to a triplet rather than a doublet will change the couplings presented below by a Clebsh-Gordan factor of order 1. In any case, for our analysis, the total widths are not important:
the $\tilde{b}_R$, $\tilde{b}_L$ and $\tilde{q}$ decay almost as soon as
they are produced.

In our simulation, we have chosen to illustrate the signal in the case where $t_R$ belongs to {\bf (1,3)+(3,1)}. The number of $4W$ events at the end will thus be  $2(1+B^2)/B^2$ times the number of events obtained from just one $\tilde{b}_R$ decaying to $tW$ (as given in Fig.~\ref{Events}), where $B$ is the relevant branching ratio.

To summarize this discussion,  whatever model is being considered, there will be at least one light KK quark, related to the SM top quark, that will lead to our $4W$ signature.
Additional light KK quarks will also contribute.
Note also that the value of the Higgs mass  depends in principle on the choice of embedding for the top and bottom quarks. In Ref.~ \cite{Contino:2006qr,Carena:2006bn},   $m_H$ and $m_{\tilde{b}_R}$ are strongly related. Models with gauge-Higgs unification typically have a light Higgs, while  for the present  analysis we allow the Higgs mass to be free and as heavy as 300 GeV.

\section{Preferred decays and event selection}
\label{eventselection}

In this study, we
focus on those decay channels which show promise for being detected
in the $pp$ collisions of the LHC.  First, we rely on a single isolated
lepton arising from one of the $W$'s to provide a clean and efficient
trigger.  We then reconstruct other $W$'s in the event using pairs of
jets.  $W$ bosons mostly decay into a pair of quarks ($BR=67.96\%$),
while they decay into a charged and neutral lepton pair about 10.68\%
of the time for each lepton generation.

The actual rate of multi-$W$ events arising from our signal depends
upon the $\tilde{b}_R$ mass as well as the Higgs mass:
for instance, in the case of $m_{\tilde{b}_R} \lsim 500$ GeV, the
branching ratio into $bZ$ increases dramatically from the $\sim 20\%$
characteristic of higher $\tilde{b}_R$ masses, leading to decay
signatures such as 2 $b$ jets + missing $E_T$, 2 $b$ jets + 2 leptons +
missing $E_T$, or 2 $b$ jets + 4 leptons.  We do not pursue
these signatures here.

For $m_{\tilde{b}_R} \gsim 500$ GeV, however, the Higgs mass plays an
important role.  If $m_H\sim 115$ GeV, it decays mainly into $b$ quarks,
and the $\tilde{b}_R\rightarrow bH$ channel leads to 6 $b$ jets.
Since the branching ratio for $bH$ is typically $\sim 50\%$, this takes
half of the decays away.  On the other hand, if the Higgs decays mainly
into $WW$, then even the $\tilde{b}_R$ decay into $bH$ produces the
4 $W$ + 2 $b$ signature.  In this case, most of the produced $\tilde{b}_R$
lead to this signature.
Note that we are assuming
that $\tilde{b}_R$ is the lightest of the KK fermions, as this is the natural
situation in minimal models. However, it could happen, such as in the GUT models
of Ref.~\cite{Agashe:2004ci,Agashe:2004bm}, that there are lighter KK states
such as the KK RH neutrino. The Higgs would then mainly decay into this KK RH
neutrino, the $H \rightarrow WW$ branching ratio would practically vanish, and
we would  lose the possibility of increasing the $WW$ yield from the Higgses.

If the Higgs mass is in the region where the dominant decay is actually to
$WW^*$, the visible signal rate will also be reduced, since $W^*\rightarrow jj$
will not peak very much around the $W$ mass.  Of course, if
$W^*\rightarrow l^{-} \nu$ and $W\rightarrow jj$, we  still get our trigger and
our $W$ combination.  For $m_H>300$ GeV, $H \rightarrow WW \rightarrow l \nu
jj$ becomes significant, so $m_H=300$  GeV is a suitable choice for a first
look at multi-$W$ signatures. The branching fraction of $H \rightarrow WW$ is
about 70\% in the $m_H$ region of 200 to 350 GeV, where the $t\bar{t}$
channel opens up. 

It should also be noted that in a typical detector, it is not trivial
to measure the charge of a jet's precursor; moreover, the jet energy is
reconstructed with a finite resolution.  For instance,
for dijet pairs with $p_T>350$ GeV, the ATLAS Technical Design Report 
\cite{ATLAS:TDR} claims
a mass resolution of 6.9 GeV, which is not much less than the
difference between the $W$ and $Z$ masses.  The two mass peaks will
overlap significantly, and indeed may be seen as a single bump.
The proposed detection method will therefore have difficulty
differentiating genuinely $WWWW$ events from, for instance, $WWZZ$
arising from a $\tilde{b}_R$ decaying to $bH$, followed by
$H\rightarrow ZZ$.  For the purpose of discovering $\tilde{b}_R$,
it is not clear that there is any advantage in discriminating
between $W$'s and $Z$'s here.  In the present signal simulation,
we do not consider these possibilities, but in principle, we should
simulate these other decays as they can potentially increase our signal.

\subsection{Backgrounds}

The Standard Model background for 4$W$ production 
(see Ref.~\cite{Barger:1989cp,Barger:1991jn}) is suppressed by high powers
of electroweak couplings and we neglect it.
The dominant background is actually fake and
comes from the misinterpretation of jets as coming from $W$'s. The two
important backgrounds we consider arise from $t \overline{t}$ and
$t\overline{t}H$ production.  $t \overline{t}$ leads to 2 $W$'s and 2 $b$'s
with four extra jets misinterpreted as coming from hadronic decays of $W$'s.
At large Higgs mass, where the Higgs decays primarily into $WW$, $t\overline{t}H$
exactly leads to $4W+2$-$b$-jets and becomes a serious background for
heavy $\tilde{b}_R$ masses:  if $m_{\tilde{b}_R}\sim 1\;{\rm TeV}$, 
the two production cross sections are comparable \cite{Dawson:2002tg}.
Distinguishing the signal in this case would be a significant challenge.




Diboson EW production can also contribute by misidentification and event
overlaps.  However, the LHC production cross sections for these processes are
smaller with respect to $t\bar{t}$ production. In addition, they do not produce
a pair of high $p_t$ b-jet pairs. Therefore, contributions  from such processes
are not currently estimated.  A list of diboson production cross sections taken
from Ref.~\cite{Campbell:1999ah} are listed below along with that of
$t\overline{t}$ as taken from the ATLAS Technical Design Report and
references therein~\cite{ATLAS:TDR,Bonciani:1998vc}:

\begin{center}
\begin{tabular}{ll}
$t\bar{t}$& 833 pb (NLO + NLL resummation) \\
$W^+W^-$ & 86.7 pb \\
$WZ$ & 32.4 pb \\
$ZZ $& 12.9 pb \\
\end{tabular}
\end{center}

The cross section for $g g \rightarrow t \overline{t} WW$ is 4 orders of magnitude below that of 
$t \overline{t} $ and this background is neglected. Similarly, the $4W+4b$  background from $(t \overline{t})  (t \overline{t})$ production
can be neglected as it is more than 4 orders of magnitude smaller than the $t
\overline{t}$  cross section:  the phase space volume is scaled down by a
$1/(4 \pi)^4$ factor and there is  a $g_s^4$ suppression factor in the cross
section.  In the end, this is $\alpha_s^2/(4\pi)^2$ suppression compared to the
$t \overline{t}$ cross section.  This is consistent with the cross section
obtained with Comphep: $ 1.3 \times 10^{-2} $ pb just from the $gg$
contribution. 

Finally, there are of course Standard Model EW processes which give $W$ + $6$-jet events,
but these are of higher order, and in addition they have no $b$ jets so we ignore
them.  There are also QCD processes which will give a hard photon + $6$ jets,
with the photon converting to a high-$p_T$ lepton.
Such events should not have much
missing $E_t$ and so should also be suppressed.

\section{Signal and background simulation}
\label{sec:signal}

The Lagrangian of the model is implemented into CalcHEP
(v2.4.3\cite{calchep}), a tree level Monte
Carlo event generator that can deal with multi-particle final states. We generate $t \overline{t} WW$ events from $\tilde{b}_R$ pair production in CalcHEP, which are further processed in PYTHIA 6.4.01 \cite{Sjostrand:2006za}.
\begin{figure}[!htb]
\begin{center}
\includegraphics[height=7.5cm,width=10.3cm]{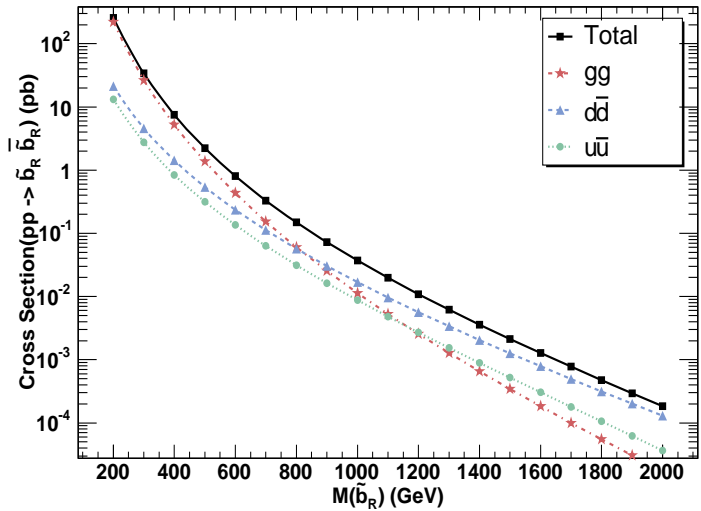}
\caption{Pair production cross section for $\tilde{b}_R$ at the LHC. 
Only masses above the $t+W$ threshold are considered.}
\label{Production}
\end{center}
\end{figure}

Figure~\ref{Production} shows the $\tilde{b}_R$ pair production cross section
at the LHC.  We use CTEQ6L parton distribution functions and consider
only the contribution from light quarks, $u$ and $d$.  The dominant
contribution comes from the gluon-initiated hard scattering up to about a
$\tilde{b}_R$ mass of 800~GeV.  The QCD scale was set to the mass of
$\tilde{b}_R$.   The cross sections include the contributions from
$s$-channel EW exchange diagrams, though the combined contribution from these channels
is negligibly small.  For example, for $\tilde{b}_R$ mass of 500~GeV, the
contribution from EW channels is about three orders of magnitude smaller than
those involving the QCD coupling.

\begin{figure}[!htb]
\begin{center}
\includegraphics[height=7.5cm,width=10.3cm]{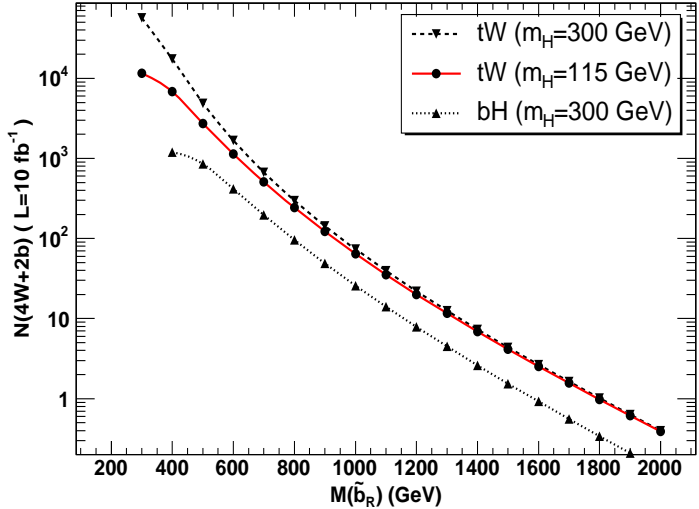}
\caption{Number of 4$W + b\overline{b}$ events expected from the $tW$ and $bH$ decays of  $\tilde{b}_R$ for 10 fb$^{-1}$ of LHC data. 
}
\label{Events}
\end{center}
\end{figure}

Figure~\ref{Events} shows the $4W+2b$ final states expected in 10fb$^{-1}$
of LHC data as a
function of the $\tilde{b}_R$ mass, for two Higgs masses and
considering only the production cross section and branching fractions.
For a $\tilde{b}_R$ mass of 500 GeV, the
yields are about 5000 events from $tW$ and 900 from $bH$ for a Higgs mass of
300 GeV. The yield is about 2800 events from only the $tW$ channel for a Higgs mass
of 115 GeV, with no contribution from the $bH$ channel.

As discussed in Section \ref{subsec:widths}, there are additional exotic quarks with exactly the same properties as $\tilde{b}_R$ which therefore enhance the total number of events:  for $m_{\tilde{b}_R}=500\;{\rm GeV}$ and
$m_H=300\;{\rm GeV}$, including $\tilde{b}_{R,L}$ and $\tilde{q}_{R,L}$,
this multiplying factor amounts to 11.2.
In the case of the models discussed in Ref.~\cite{Contino:2006qr,Carena:2006bn,Djouadi:2006rk,Cacciapaglia:2006mz}, where some have only one light $Q=5/3$ quark $\tilde{q}$ instead of
$\tilde{b}_R$, this factor is $1/B^2=4.6$.

\subsection{Trigger strategy and event reconstruction}


In order to investigate how such events could be examined using an
LHC detector such as ATLAS, we first restrict ourselves to
simulating $\tilde{b}_R$ pair production and their decays through
the $tW$ channel. For the present model analysis, we choose $m_{\tilde{b}_R}=500$ GeV
and $m_H=300$ GeV.
We apply event filters and acceptance
criteria based on nominal ATLAS parameters as found in the Technical
Design Report~\cite{ATLAS:TDR}.  The following ``trigger'',
applied to the generated events, is based on the lepton criteria
for selecting $W\rightarrow\ell\nu$ events:  at least one electron or
muon with $p_T>25\;{\rm GeV}$ must be found within      the pseudorapidity
range $|\eta|<2.4$, where $\eta = -\ln\tan(\theta/2)$ and $\theta$ is the angle
relative to the $pp$ collision axis; then, the ``missing $E_T$'', calculated by
adding all the neutrino momenta in the event and taking the component transverse to
the collision axis, must exceed 20 GeV.

\begin{figure}[tbph]
\begin{center}
\includegraphics[width=10.3cm]{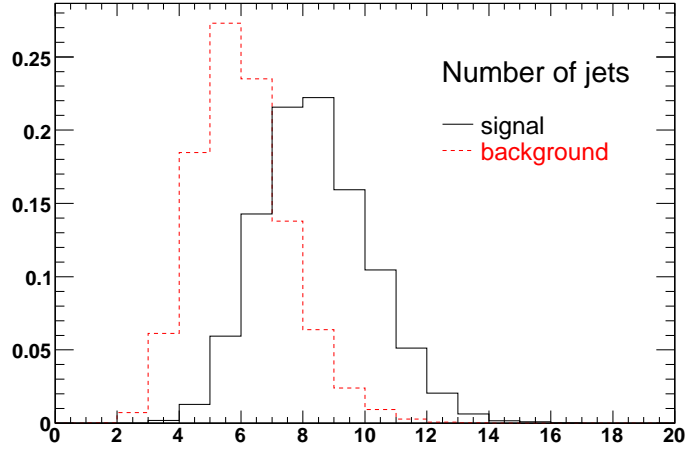}
\end{center}
\caption{Number of jets with $p_T > 20\;{\rm GeV}$ for signal and
background.  Both distributions
are normalized to unit area.}
\label{njet}
\end{figure}

We mimic cone-based hadronic jets as they might be observed in
a detector:
stable charged and neutral particles within
$|\eta|<4.9$ (the range of the ATLAS hadronic calorimeter),
excluding neutrinos, are first ranked in $p_T$ order.  Jets are seeded
starting with the highest $p_T$ tracks, with $p_T>1\;{\rm GeV}$;
softer tracks are added to the
nearest existing jet, as long as they are within $\Delta R < 0.4$
of the jet centroid, where $\Delta R = \sqrt{\Delta\phi^2 + \Delta\eta^2}$.
The number of jets with $p_T > 20\;{\rm GeV}$ is shown in Figure~\ref{njet}.
The signal is peaked around 8 jets.  Additional jets can be produced
in quark hadronization and underlying parton activity, while jets can
be lost due to falling below the energy threshold or outside the
geometric acceptance.

The background sample is dominated by $t\overline{t}$ events generated
using TopRex (version 4.11) \cite{Slabospitsky:2002ag} and PYTHIA (version 6.403), with CTEQ6L parton
distribution functions.  The small $t\overline{t}H$ contribution to
the background has been modelled with PYTHIA.  As expected,
the background has fewer high-$p_T$ jets than
the signal, peaking around 5 jets.

\begin{figure}[tbph]
\begin{center}
\includegraphics[width=8.3cm]{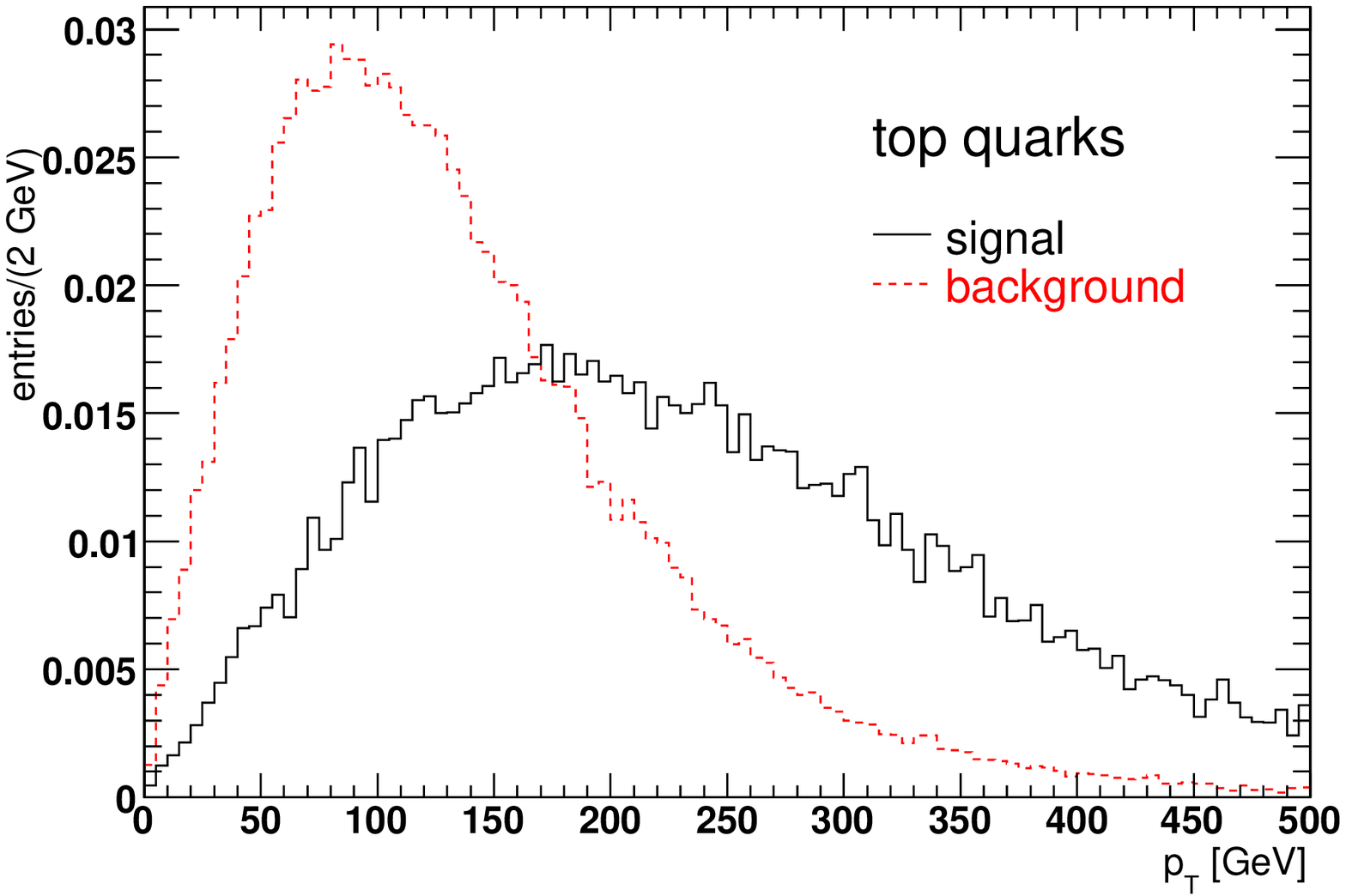}
\includegraphics[width=8.3cm]{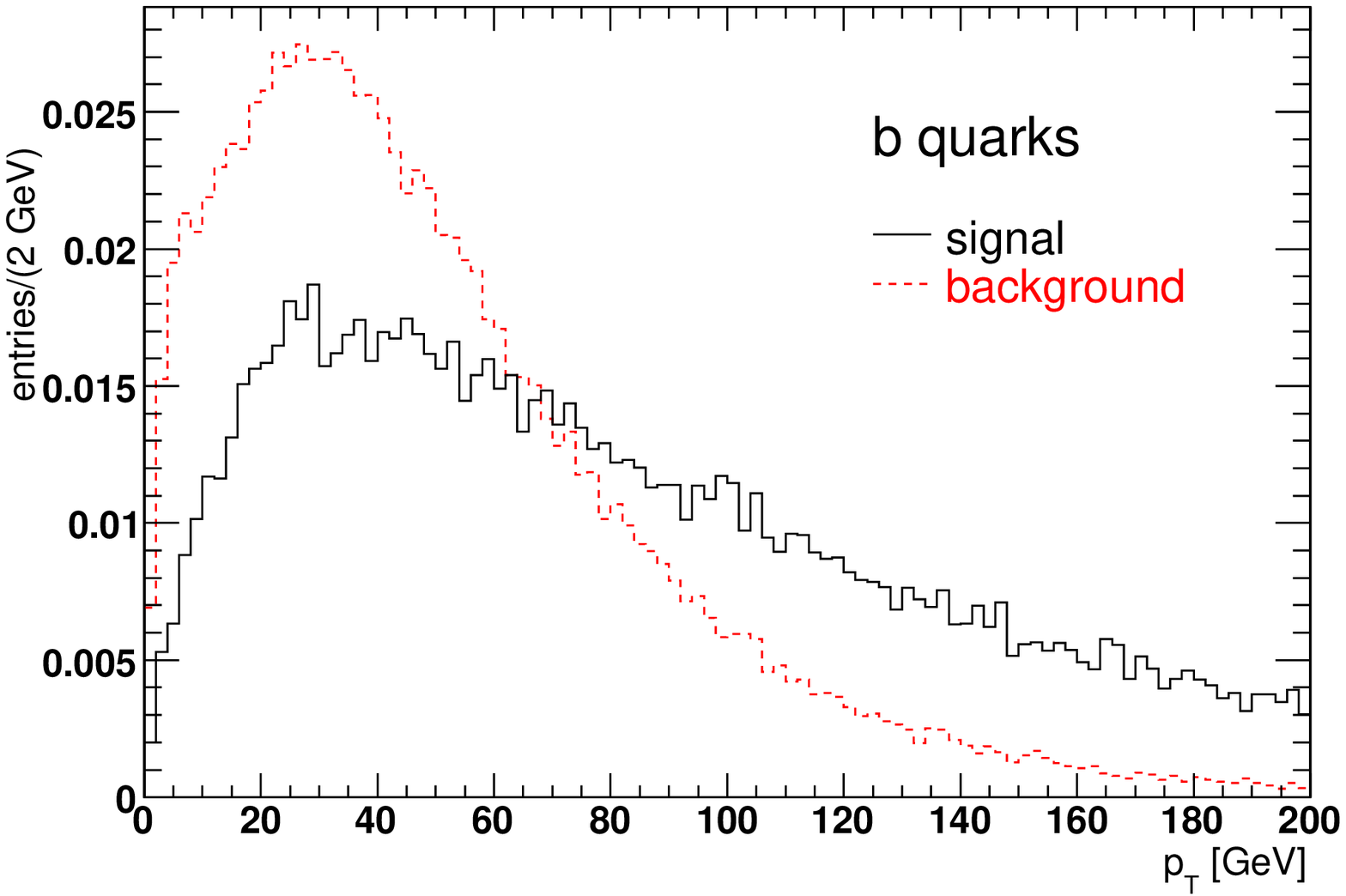}
\includegraphics[width=8.3cm]{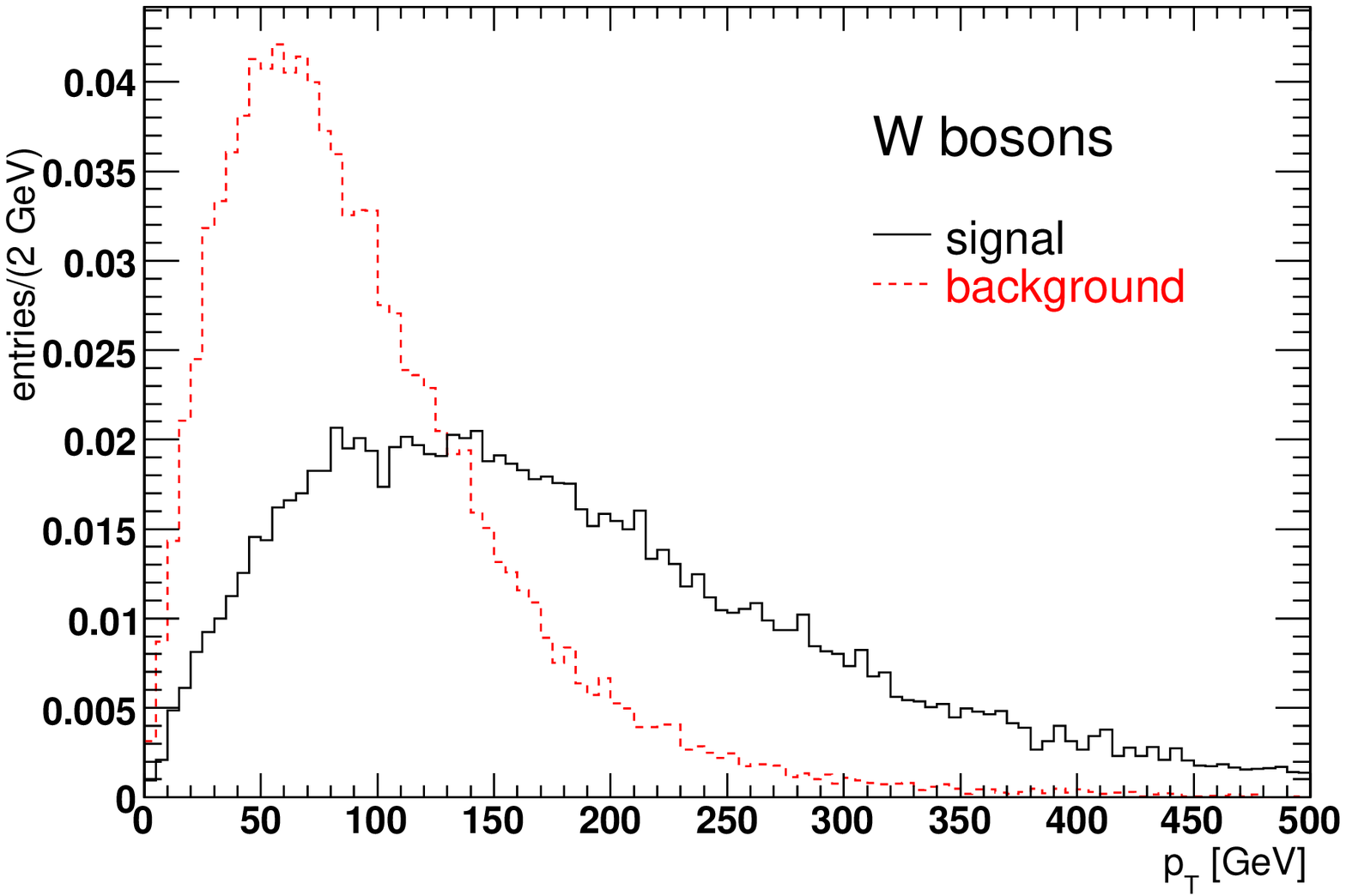}
\includegraphics[width=8.3cm]{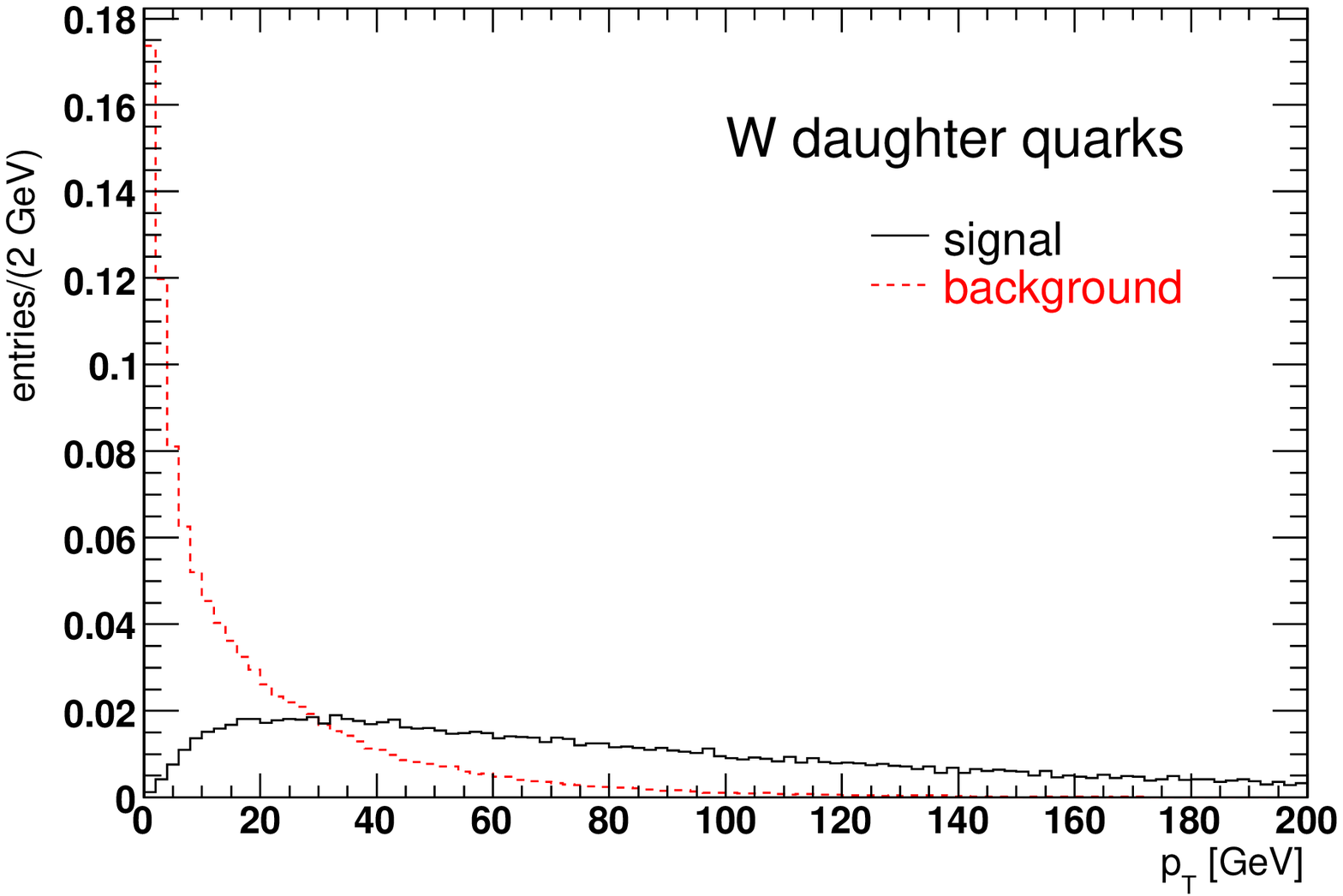}
\end{center}
\caption{Generator-level signal and background $p_T$ distributions after the
``trigger'' conditions, normalized to unit
area.  Top left:  $t$ quarks.  Top right:  $b$ quarks.  Bottom left:
$W$ bosons.  Bottom right:  $W$ daughter quarks.}
\label{genpt}
\end{figure}

\begin{figure}[tbph]
\begin{center}
\includegraphics[width=8.3cm]{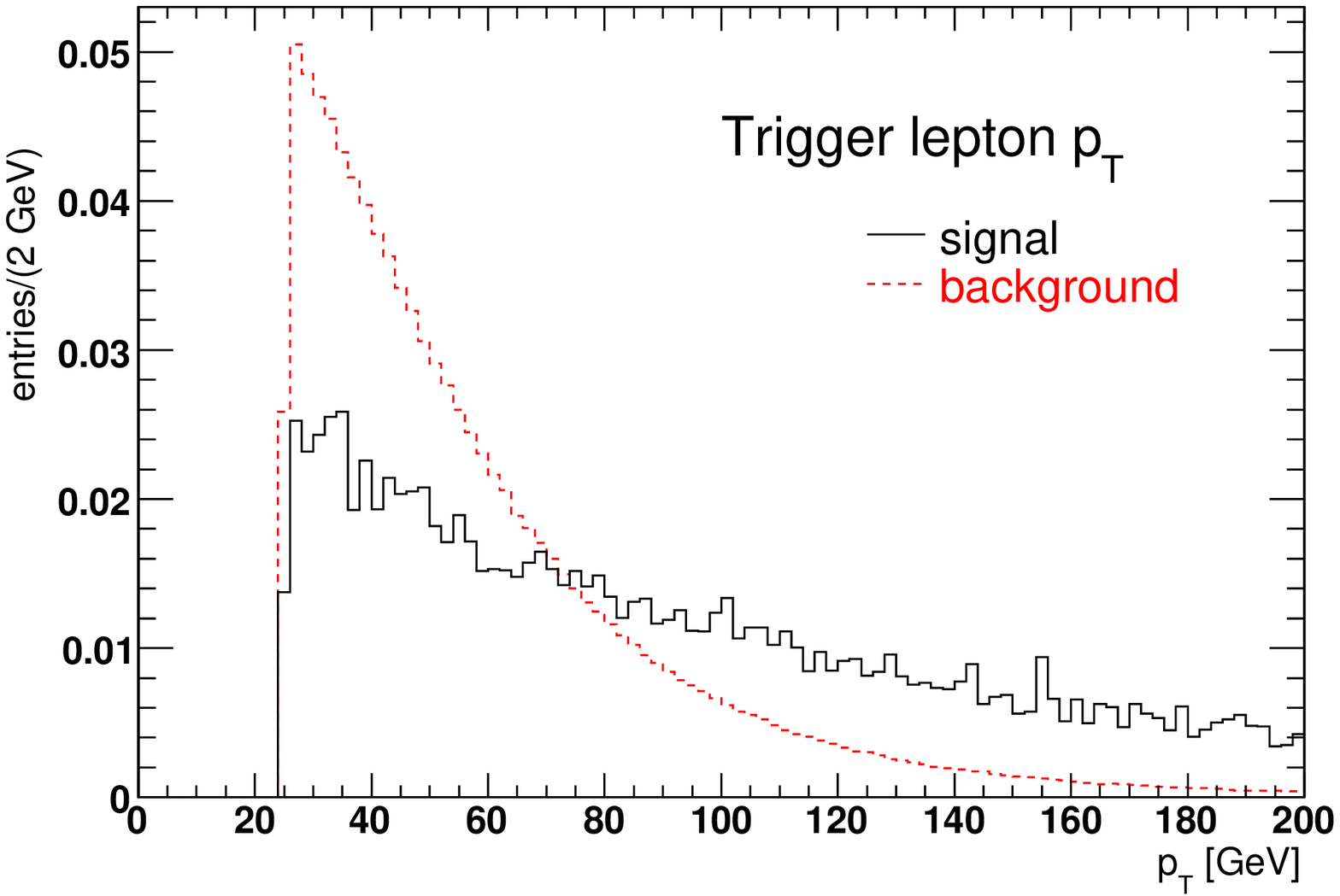}
\includegraphics[width=8.3cm]{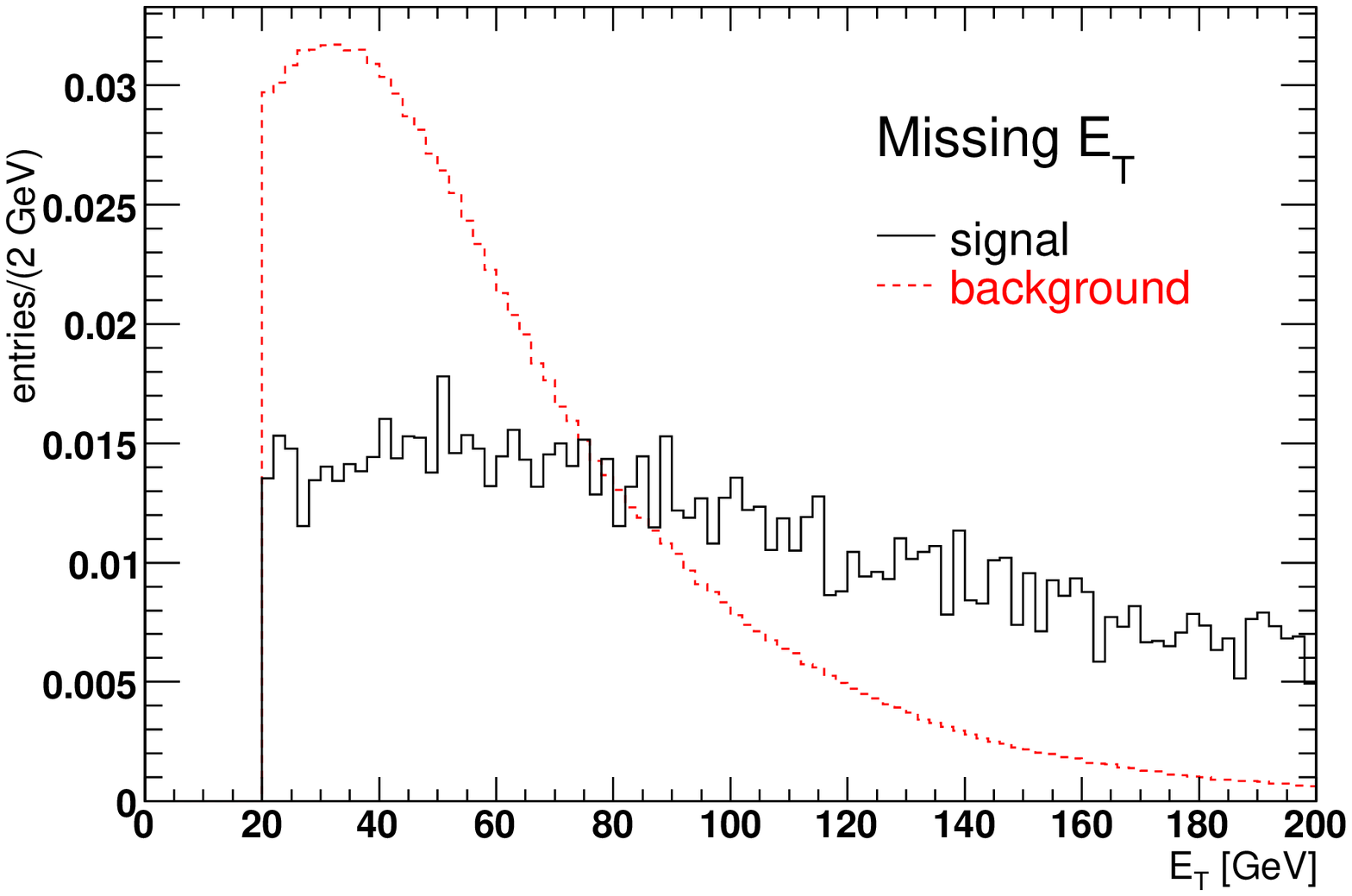}
\includegraphics[width=8.3cm]{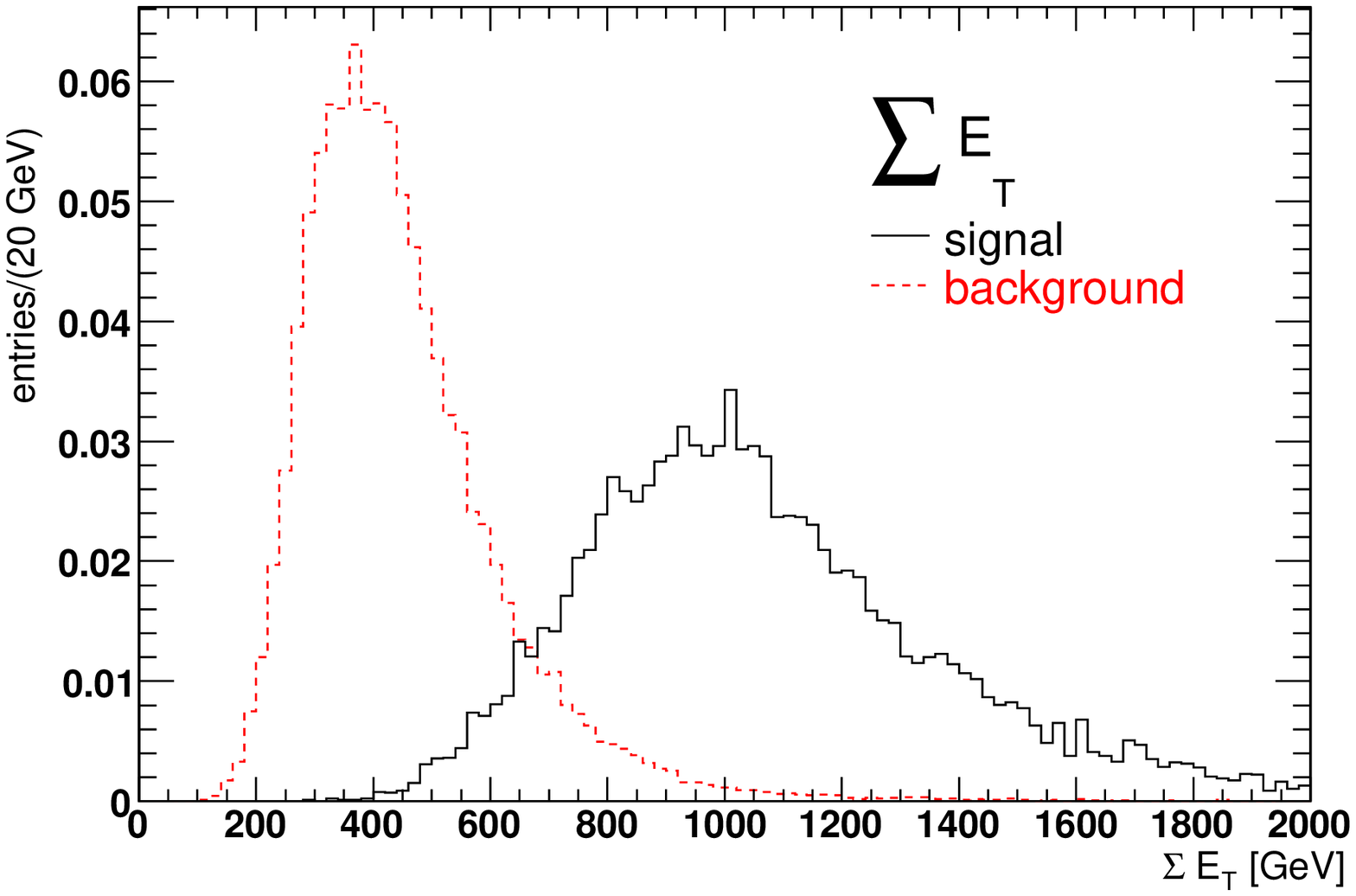}
\includegraphics[width=8.3cm]{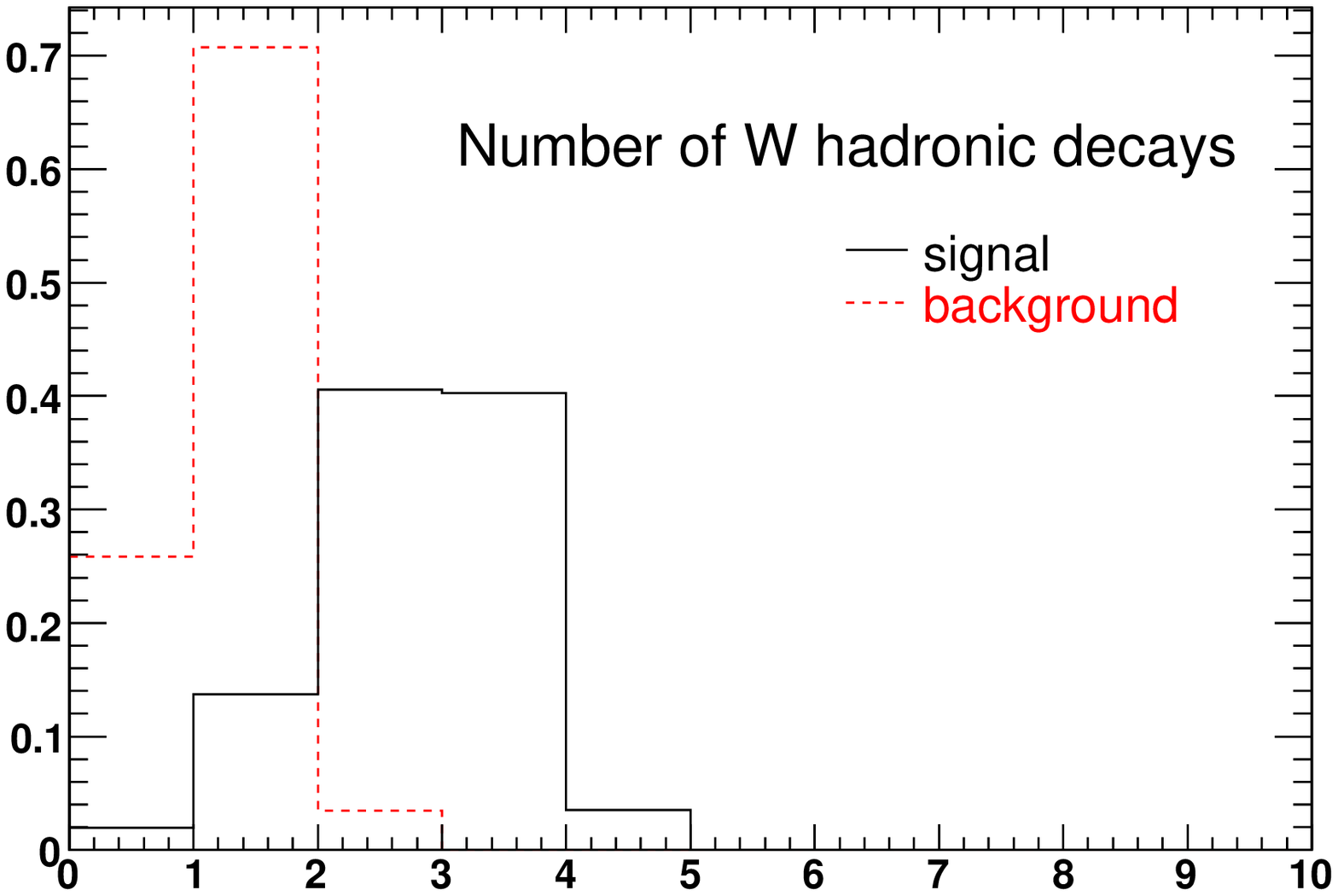}
\end{center}
\caption{Generator-level signal and background ``trigger'' distributions after 
initial trigger conditions, normalized to unit
area.  Top left:  $p_T$ of stiffest electron or muon.  Top right:  missing
$E_T$.  Bottom left:  scalar sum of $E_T$ in the event.  Bottom right:
number of $W$'s decaying hadronically in the event.}
\label{trigpt}
\end{figure}

Figures~\ref{genpt} and \ref{trigpt} compare signal and background
distributions for transverse momenta of generated particles as well as
event-level observables such as the ``trigger'' (highest-$p_T$) lepton
$p_T$, missing $E_T$, and scalar sum of the $E_T$'s of all the jets
in the event.

The scalar $E_T$ sum distribution in Figure~\ref{trigpt} (bottom left)
suggests that the background
can be significantly reduced by requiring the sum to
exceed 800 GeV.  We also restrict the analysis to
events with between 6 and 9 jets.
No $b$ jet tagging is performed at this point.



\begin{figure}[tbph]
\begin{center}
\includegraphics[width=8.3cm]{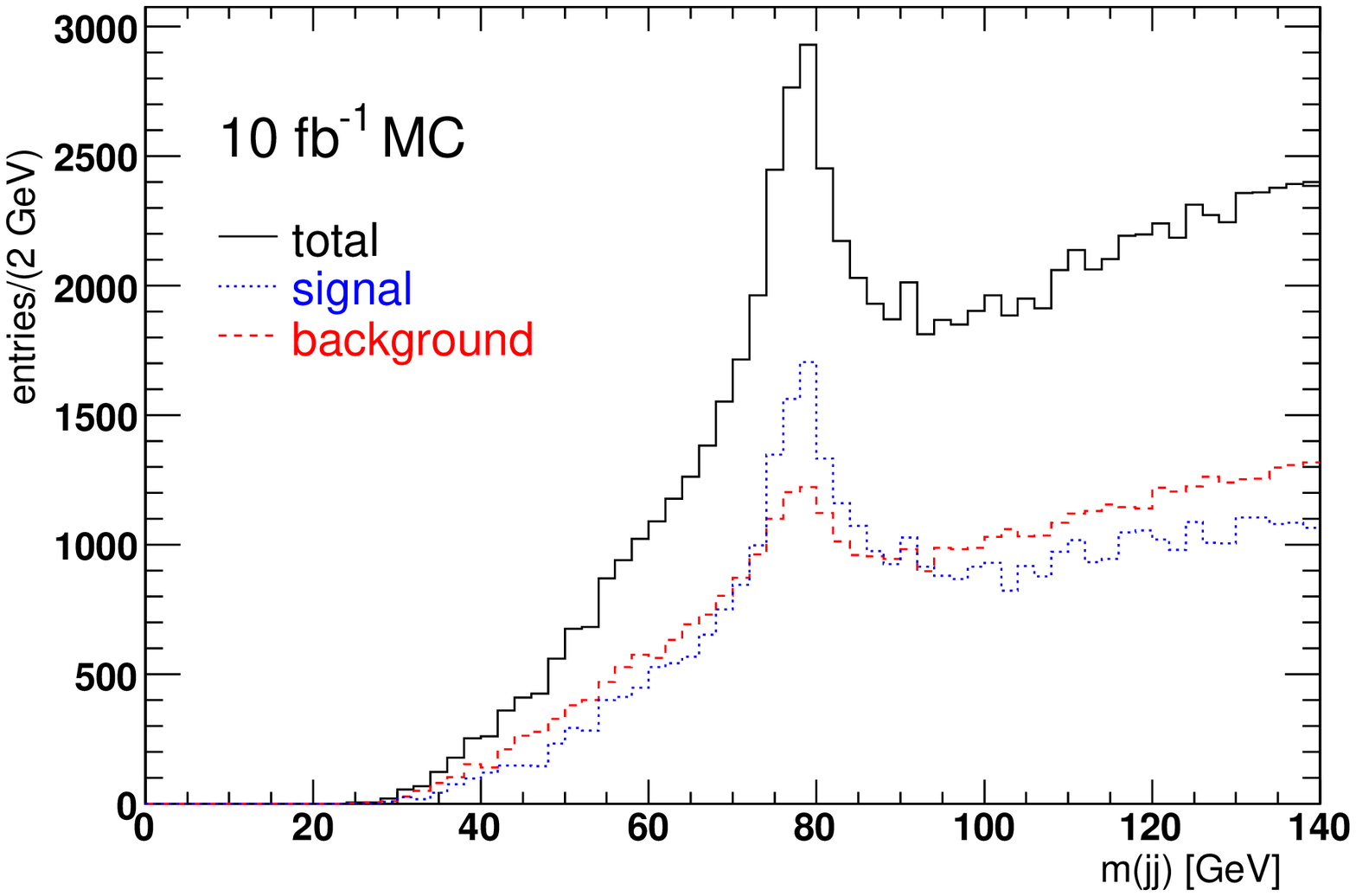}
\includegraphics[width=8.3cm]{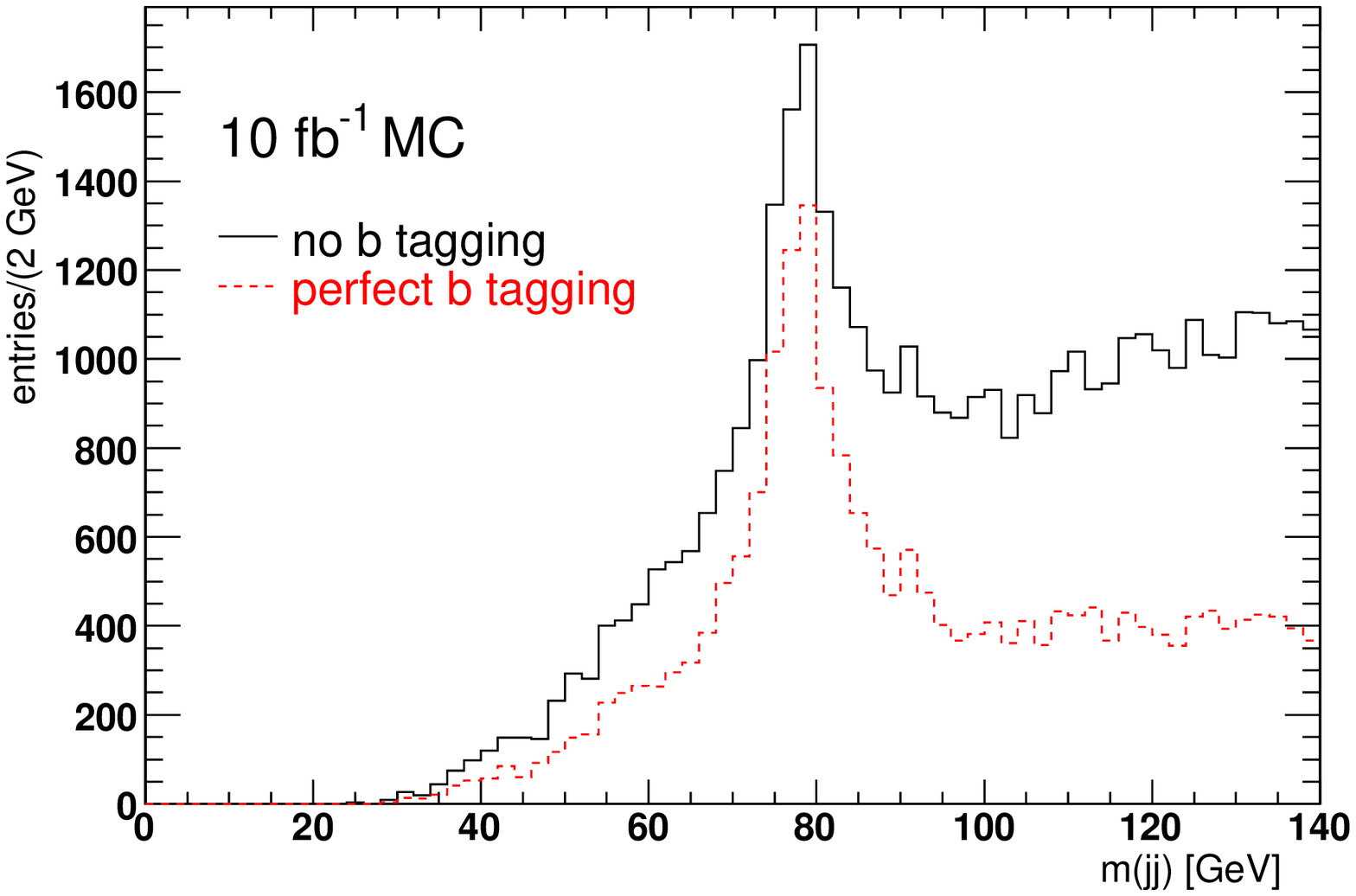}
\end{center}
\caption{Left:  dijet mass distribution for all jets for signal,
background, and their sum, for $10\;{\rm fb}^{-1}$ integrated
luminosity.  Right:  comparison of
dijet mass distributions before and after eliminating all
$b$ jets.  The black histogram here is the same as the blue in the
figure to the left.}
\label{dijetmass}
\end{figure}

After applying the 800 GeV cut to the scalar $E_T$ sum,
we construct
hadronic $W$ candidates by adding the 4-momenta of two jets,
each with $p_T>20\;{\rm GeV}$ and assuming that each individual jet
has zero mass.  Since the $W$'s are typically produced with
$p_T > 150\;{\rm GeV}$, as shown in Figure~\ref{genpt},
the total $p_T$ of the dijet combination is further
required to exceed that value.  The mass spectrum for the signal and
background events,
representing $10\;{\rm fb}^{-1}$ of LHC data,
is shown in Figure~\ref{dijetmass}.
We have scaled up the signal to include the $\tilde{b}_L$ and $\tilde{q}$
relatives of the $\tilde{b}_R$, as discussed in Section~\ref{subsec:widths}
(this scaling assumes charge-symmetric lepton and jet identification).
A prominent peak can be seen around 80 GeV with width 5 GeV.
The width is due to, among other effects, the hadronization of the
daughter quarks, with the resulting jets sometimes overlapping with
other activity in the event.  The low mass of the peak, relative to
the generated $W$ mass, is expected given, for instance, the finite
cone size of the jets.  A $W$ peak is also evident in the
$t\overline{t}$-dominated background distribution, as expected.
Figure~\ref{dijetmass} also shows the reduction in the signal's
combinatorial
background if perfect $b$ tagging is achieved, though experimentally,
$b$ tagging is rarely very efficient or pure.

\begin{figure}[tbph]
\begin{center}
\includegraphics[width=10.3cm]{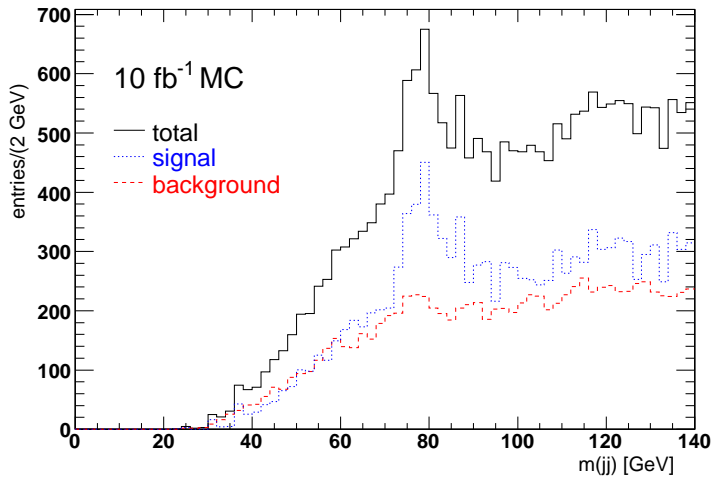}
\end{center}
\caption{Dijet mass distribution and signal, background,
and their sum,
after eliminating the first hadronic $W$ candidate.  The distribution
represents $10\;{\rm fb}^{-1}$ integrated luminosity.}
\label{rank1}
\end{figure}

In order to suppress the most common ($t\overline{t}$) Standard Model
background, we eliminate the single hadronic $W$ decay in the following
manner:  we start with the highest $p_T$ jet and search among the
lower $p_T$ jets for a combination whose mass falls between 70 and 90 GeV.
If no combination is found, the search is continued using the next
highest $p_T$ jet.  If a pair is found, those two jets are removed for
consideration in forming the subsequent dijet mass combinations, which
are shown in Figure~\ref{rank1}.  The observable peak is by now
dominated by the signal, though the background peak has
not been entirely eliminated.  These background events might appear,
paradoxically, to contain 3 $W$'s, one decaying leptonically and two
hadronically, but a more mundane explanation is that the trigger
lepton and missing energy actually arise from semileptonic $b$ decay
occurring among the far more numerous $t\overline{t}$ events.  These
events also appear in Figure~\ref{trigpt} (bottom right) as
background events with 2 hadronic $W$ decays; a similar fraction
of signal events contain 4 hadronic $W$ decays.  While the
background peak may not be very large compared to the signal, it
can still be further suppressed by increasing the trigger thresholds.

Relating the $\tilde{b}_R$ production cross section to the
size of the peak (and therefore the discovery potential of this
signature) is non-trivial:  first,
we would have to subtract or suppress the $t\overline{t}$
component, model the dijet mass resolution, and calculate the
reconstruction efficiency in light of the multiple $W$ decays in
the event.
Such details are best reserved for further investigation,
when a more detailed and specific simulation of the detector and
experimental environment, including the effects
of detector material and
multiple $pp$
interactions in a beam crossing, can be employed.
Furthermore, the relationship between $\tilde{b}_R$ production
and yield is complicated by the $m_H$-dependent interplay of
different decay modes and how they contribute $W$'s and
$Z$'s to the signal.  However, this initial analysis, indicative
of the kind which could be attempted at an LHC experiment,
suggests that it will be possible to distinguish the signal from
the major physics backgrounds.

\subsection{Further analysis directions}

\begin{figure}[tbph]
\begin{center}
\includegraphics[height=6.25cm, width=10.3cm]{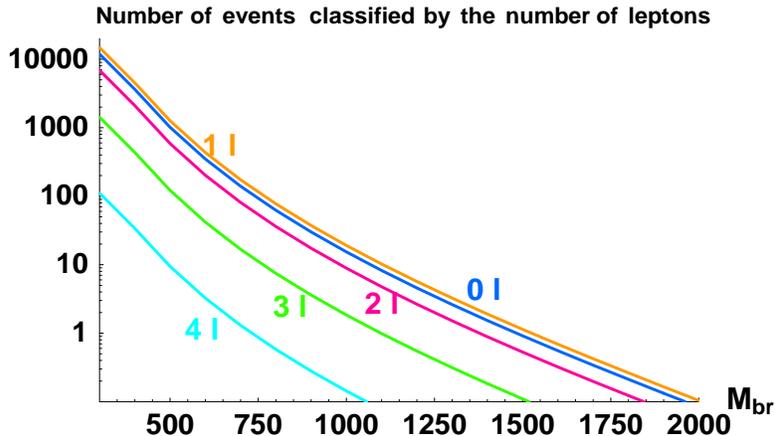}
\end{center}
\caption{Number of $4W$ events (coming from $\tilde{b}_R$ only) in the case $m_H=300$ GeV, with the
assumption of $10\;{\rm fb}^{-1}$ integrated luminosity. Leptons are either $e$ or $\mu$.}
\label{leptons}
\end{figure}

It has been suggested in Ref.~\cite{Carena:2006bn} that if all four $W$'s decay leptonically, the
resulting ``golden'' signature of 4 leptons + $b\overline{b}$ + missing $E_T$
would have very little SM background.  As shown in Figure~\ref{leptons},
however, the rate for this signature would be suppressed by approximately
two orders of magnitude relative to a signature requiring only one
leptonic $W$ decay.  At the same time, because of the multiple neutrinos,
it would be difficult to identify the leptons as $W$ daughters.  As a result,
while this signature could be indicative of new physics, it would be
difficult to distinguish what kind of new physics is being observed.

%

An important piece of corroborative evidence for these models is the
observation of the custodian quark $\tilde{q}$ carrying electric
charge $Q=5/3$.  Pair production of this quark would lead to exactly
the same 4 $W$ + $b\overline{b}$ final state as the $\tilde{b}_R$.
A possible method to distinguish the $\tilde{q}$ events is to
select events with two high-$p_T$ leptons with the same sign
(presumably from the two $W$'s from a single $\tilde{q}$), and
then fully reconstruct the other $\tilde{q}$ in the event through
its $tW\rightarrow bWW\rightarrow bjjjj$ decay.
This method is driven by the typical situation that the charge of an
observed lepton can be measured reliably, while that for a jet cannot.
The fully reconstructed $\tilde{q}$ would yield a narrow peak in the
$b$ + 4 jet combined invariant mass distribution, while the
corresponding distribution arising from $\tilde{b}_R$ pair production
and decay would be much broader.  
The main disadvantage of this method is the typically low efficiency
of a full reconstruction, which requires all decay products to be
observed and well measured.  We therefore expect that confirming the
existence of the $\tilde{q}$ custodian would require rather more data
than the $10\;{\rm fb}^{-1}$ studied here (note, however, that the
LHC's design luminosity would yield $100\;{\rm fb}^{-1}$ per year).

Finally, 
let us note that one could also consider the single production
of $\tilde{b}_R$ together  with the standard  $b$ quark via s-channel $Z$ or  $W$-gluon fusion, or together with the top quark via s-channel $W$. 
This would be especially relevant for large $\tilde{b}_R$ masses and would deserve a
separate study.

\section{Conclusion and future prospects}

We have studied new signals in pair production of heavy $Q=-1/3$ and $Q=5/3$ quarks
at the LHC. They are produced through standard QCD
interactions with a cross section $\sim {\cal O}(10)$ pb for masses of several
hundreds of GeV.  Heavy quarks such as $\tilde{b}_R$ are well-motivated in
Randall--Sundrum models with custodial symmetry. They are generally Kaluza-Klein partners of the Standard Model Right-Handed top quark. Their decay channels were
described in detail in this paper.  In the present work, we focussed on the
4-$W$ events which we believe are quite specific to this class of models, and
also experimentally promising. We have considered the process $gg, q
\overline{q}\rightarrow  \tilde{b}_R  \overline{\tilde{b}}_R \rightarrow W^- t
\ W^+  \overline{t}  \rightarrow W^- W^+ b \ W^+  W^- \overline{b}$ where at least one
$W$ boson decays leptonically and the other ones hadronically. A simulation
of this signal and its main background was performed, and an analysis
strategy outlined which distinguishes the signal from the sizable
Standard Model backgrounds.
The peak we obtain in the dijet mass distribution suggests that it is possible to reach a signal significance beyond the $5\sigma$ level. Further study with more detailed simulation
is required to map the
discovery potential for this signal at an LHC experiment such as ATLAS,
or at the ILC, and to connect the observable signal to the production
cross section.

\section*{Acknowledgments}

We acknowledge the financial support of the Particle Physics and
Astronomy Research Council of the United Kingdom.
We are indebted to Kaustubh Agashe for collaboration and many helpful
discussions.  We also  thank Sasha Pukhov and Genevi\`eve B\'elanger for
checking the Comphep file and especially S. Pukhov for valuable help on CalcHEP
related questions.  G.S is grateful to Henry Frisch and  Drew Fustin
for collaboration at some very early stage of this project. 
Finally, G.S thanks Roberto Contino for useful comments.


\end{document}